\documentclass[namedreferences]{solarphysics}

\usepackage{graphicx}        
\usepackage{float}

\usepackage{longtable}
\usepackage{amssymb}        
\usepackage{color}           
\usepackage{breakurl}        
\usepackage{array, booktabs}
\usepackage{bm}
\usepackage[hyperref,optionalrh]{spr-sola-addons} 

\chardef\us=`\_

\begin{document}

\begin{article}
\begin{opening}

\title{Identifying Flux Rope Signatures Using a Deep Neural Network \\ {\it Solar Physics}}

\author[addressref={aff1,aff2},corref,email={luizfernando.guedesdossantos@nasa.gov}]{\inits{L. F.}\fnm{Luiz F.}~\lnm{G. dos Santos}\orcid{0000-0001-5190-442X}}
\author[addressref={aff1,aff3},email={ayris.a.narock@nasa.gov}]{\inits{A.}\fnm{Ayris}~\lnm{Narock}\orcid{0000-0001-6746-7455}}
\author[addressref={aff1}]{\inits{T.}\fnm{Teresa}~\lnm{Nieves-Chinchilla}\orcid{0000-0003-0565-4890}}
\author[addressref={aff4},]{\inits{M.}\fnm{Marlon}~\lnm{Nuñez}\orcid{0000-0001-5374-5231
}}
\author[addressref={aff1,aff5},]{\inits{M.}\fnm{Michael}~\lnm{Kirk}\orcid{0000-0001-9874-1429}}


\address[id=aff1]{Heliophysics Science Division, NASA, Goddard Space Flight Center, Greenbelt, MD 20771, USA}
\address[id=aff2]{The Catholic University of America, 620 Michigan Avenue NE, Washington, DC 20064, USA}
\address[id=aff3]{ADNET Systems Inc., 7515 Mission Drive, Lanham, MD 20706, USA}
\address[id=aff4]{Departamento de Lenguajes y Ciencias de la Computación, Universidad de Málaga, Campus de Teatinos, 29071 Málaga, Spain}
\address[id=aff5]{ASTRA, LLC., 282 Century Place Suite 1000, Louisville, CO 80027, USA}

\runningauthor{L. F. G. dos Santos et al.}
\runningtitle{Identifying Flux Rope Signatures Using a Deep Neural Network}

\begin{abstract}
Among the current challenges in Space Weather, one of the main ones is to forecast the internal magnetic configuration within Interplanetary Coronal Mass Ejections (ICMEs).  Currently, a monotonic and coherent magnetic configuration observed is associated with the result of a spacecraft crossing a large flux rope with helical magnetic field lines topology. The classification of such an arrangement is essential to predict geomagnetic disturbance. Thus, the classification relies on the assumption that the ICME's internal structure is a well organized magnetic flux rope. This paper applies machine learning and a current physical flux rope analytical model to identify and further understand the internal structures of ICMEs. We trained an image recognition artificial neural network with analytical flux rope data, generated from the range of many possible trajectories within a cylindrical (circular and elliptical cross-section) model. The trained network was then evaluated against the observed ICMEs from WIND during 1995-2015.\\

The methodology developed in this paper can classify 84\% of simple real cases correctly and has a 76\% success rate when extended to a broader set with 5\% noise applied, although it does exhibit a bias in favor of positive flux rope classification.  As a first step towards a generalizable classification and parameterization tool, these results show promise. With further tuning and refinement, our model presents a strong potential to evolve into a robust tool for identifying flux rope configurations from \textit{in situ} data.
\end{abstract}
\end{opening}

\section{Introduction}
    \label{S-Introduction} 

The main drivers of geomagnetic activity are interplanetary coronal mass ejections (ICMEs). Besides transporting large quantities of mass and magnetic flux away from the Sun, their internal magnetic field structure is often coupled to the upper magnetosphere, triggering magnetic reconnection processes that allow solar magnetic energy to be injected into the entire magnetospheric system. Thus, a reliable classification of the ICME internal magnetic field structure is a requirement to develop a robust space weather forecast. The imprints of an ICME's internal magnetic structure often present a particular type of configuration. In space weather, the classification of such an arrangement is essential to predict geomagnetic disturbance. The leading hypothesis is to assume that such a structure is a flux rope (FR). \\

Our information about the internal magnetic structure of ICMEs is limited to the 1D observations of a single spacecraft crossing the large structure, leaving a considerable amount of uncertainty about the three-dimensional structure. Magnetic Clouds (MC) \citep{Klein_Burlaga_1982, Burlaga_1981} are not always detected within the ICMEs \citep[see][]{Richard_Cane_2004} even though flux ropes are always expected based on the CME eruption theories \citep[see][and references therein]{Vourlidas_2014}. This might result from changes during interplanetary evolution \citep[see][and references therein]{Manchester_2017}, from spacecraft crossing far from the ICME core, or possibly from the topological complexity of the magnetic structure during CME initiation and evolution in the solar corona and inner heliosphere. There are many known physics-based flux rope models \citep{Leeping_Jones_Burlaga_1990} used to reconstruct the internal ICME magnetic configuration and provide information on orientation, geometry, and other magnetic parameters such as the central magnetic field. \\

Recently, \cite{Nieves_2018} carried out a comprehensive study of the internal magnetic field configurations of 337 ICMEs observed by WIND at 1 AU in the period 1995-2015 to unravel the internal magnetic structure associated with the CMEs and establish under what signatures a flux rope model is valid. The analysis adopted a less restrictive term than MC, magnetic obstacle (MO) \citep{Nieves_2018}, to describe the magnetic structure embedded in an ICME. The Magnetic Field Instrument \citep[MFI,][]{Lepping_1995} and Solar Wind Experiment \citep[SWE,][]{Ogilvie_1995} were used to set the boundaries of the MO and reconstruct those cases found with flux rope configuration. All the events were sorted into three broad categories based on the magnetic field rotation pattern: events without evident rotation (E), those with single magnetic field rotation (F), and those with more than one magnetic field rotation (Cx). Later, \cite{Nieves_2019} presented a more in-depth classification in an expanded catalog of 353 ICMEs. It further classified the F types events into $F^-$, Fr, and $F^+$ based on the angular span of the magnetic field rotation. In addition to categorizing them, the events of type F were also fitted to the Circular-Cylindrical flux rope model \citep{Nieves_C_2016}.\\

Meanwhile, the application of machine learning (ML) has also gained momentum in the space weather community \citep[see][and references therein]{Camporeale_2019}. We are observing an increase of space- and ground-based capabilities with a growing amount of data available.  Inspired by \cite{Nieves_2018} and \cite{Nieves_2019}, we take advantage of ML techniques to interpret the ICME \textit{in situ} magnetic field observations and understand in depth what \textit{in situ} magnetic field observations should be expected when a spacecraft crosses flux ropes with different trajectories. Our choice of methodology is driven by the sparse real dataset that can be represented in the form of an image but does not depend on it being a magnetic field.  There are many episodic events studied in heliophysics that may be able to apply a similar approach.\\ 

We present in this paper a demonstration of a tool using supervised learning techniques and a Deep Convolutional Neural Network (DCNN) based on handwriting recognition models to classify and analyze a subset of the events included in \cite{Nieves_2019}. In Section \ref{S-Components and Methodology}, we present our DCNN model, describe our data set, and introduce our methodology to approach the problem. Section \ref{S-results} discusses the results and Section \ref{S-Summary} summarizes the paper.

\section{Methodology}
     \label{S-Components and Methodology} 
 The novel methodology presented in this paper relies on combining artificial neural networks with our current understanding of the internal structure of the ICMEs to classify \textit{in situ} data measured by WIND spacecraft and eventually to test such knowledge. We create a machine learning approach using a Deep Convolutional Neural Network model (DCNN-model) \citep{lecun1995convolutional} and train its weights with synthetic data obtained from well established physical flux rope model. This approach is conceptually different from a more ``standard" machine learning problem in which one aims to learn about a data-space by sampling a subspace of it (e.g., to identify pictures of a cat by training on many images of cats). Afterward, we use evaluation metrics to analyze performance on a selected subset of real event data. We also use this analysis to chose the DCNN-model architecture with which we ultimately proceeded. We then added additional training and evaluation cycles using synthetic training data augmented with noise. Each extra training cycle is based on the best performing training epoch of the previous training cycle.\\
 
 In the following subsections we will present the Deep Convolutional Neural Network (Section \ref{S-machine-learning-model}), the synthetic data used for training (Section \ref{S-synthetic-data}), the real data used for evaluation (Section \ref{S-Dataset}), the methodology (Section \ref{S-Model-Evaluation}),  and the analysis of the evaluation results (Section \ref{S-real-validation}).

 \subsection{Deep Convolutional Neural Network} 
  \label{S-machine-learning-model}

Our deep convolutional neural network is a binary classification model implementation of the multi-class handwritten digit recognition models \citep{ciresan2011flexible}. The input to our model is a stack of three hodograms images (see section \ref{S-synthetic-data}), having an array dimension of (3,32,32) and the output of this model is a two-element vector describing the probability of this hodogram set being a flux rope (FR) or a Non-flux rope (NFR). \\

\begin{center}
      \begin{figure} [!ht]   
   \centerline{\includegraphics[width=1\textwidth,clip=]{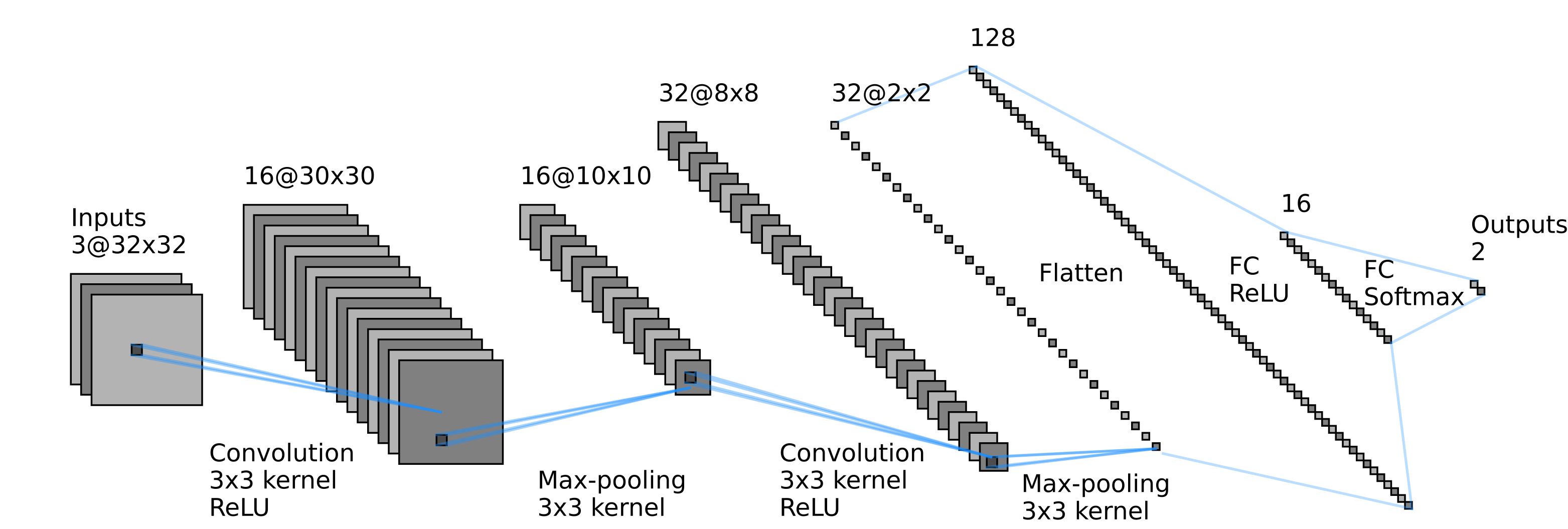}
              }
              \caption{Schematic of the DCNN architecture used for the model.The blue arrows are the architecture layers and grey squares represent the output of each layer. The architecture consists of two sets of convolutional and max-pooling layers using ReLU activation, followed by two fully connected layers with ReLU and softmax as the activation functions, respectively. All convolution and max-pooling layers have a kernel size of $(3,3)$.}
   \label{F-architecture}
   \end{figure}
\end{center}

Figure \ref{F-architecture} shows a schematic of the DCNN architecture used for our DCNN-model. The gray squares represent how the input hodogram, with shape (3, 32, 32), is changing after each layer of the DCNN. From left to right we have a convolution layer, ReLU \citep[Rectified Linear Units][]{Nair_2010} activation,  max-pooling layer, convolution layer, ReLU activation, Max Pooling layer, Flatten layer, a Fully Connected layer with 128 inputs and 16 outputs, a ReLU activation, and a Fully Connected layer with 16 inputs and two outputs and a Softmax activation. All convolution and max-pooling layers have a kernel size of (3,3). The model and training were implemented with PyTorch \citep{paszke_2017} version 1.3.1 in a Python 3.6.8 environment.

 \subsection{Synthetic Data} 
  \label{S-synthetic-data}
  
The DCNN network weights are trained using synthetic data created from two different sources: flux ropes (FR) from a physics-based model and non-flux ropes (NFR) from a empirical model developed for this work.\\

\begin{figure}[ht]
\centerline{\textbf{a}\includegraphics[height=0.45\textwidth,clip=]{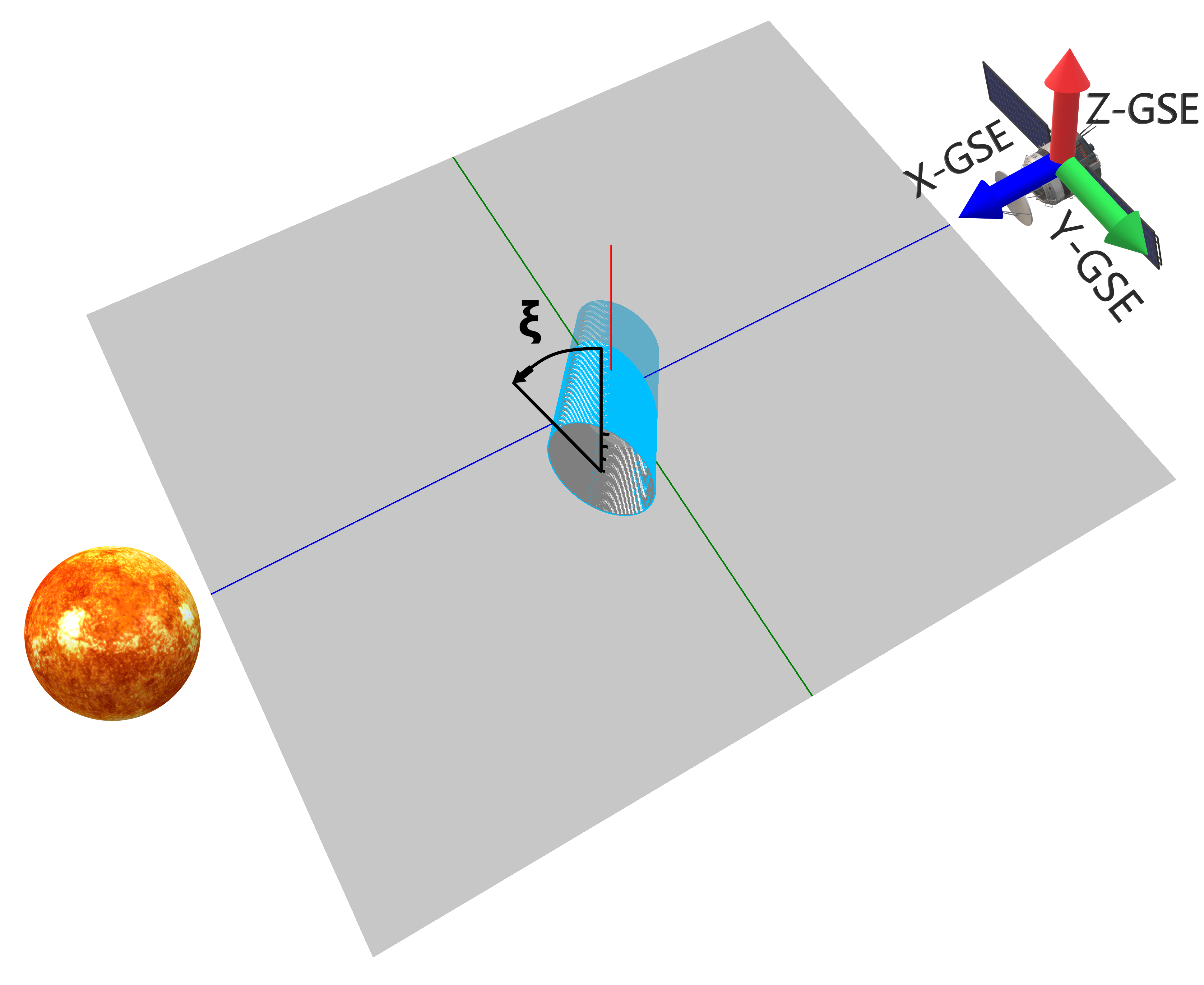}
            \textbf{b}\includegraphics[height=0.45\textwidth,clip=]{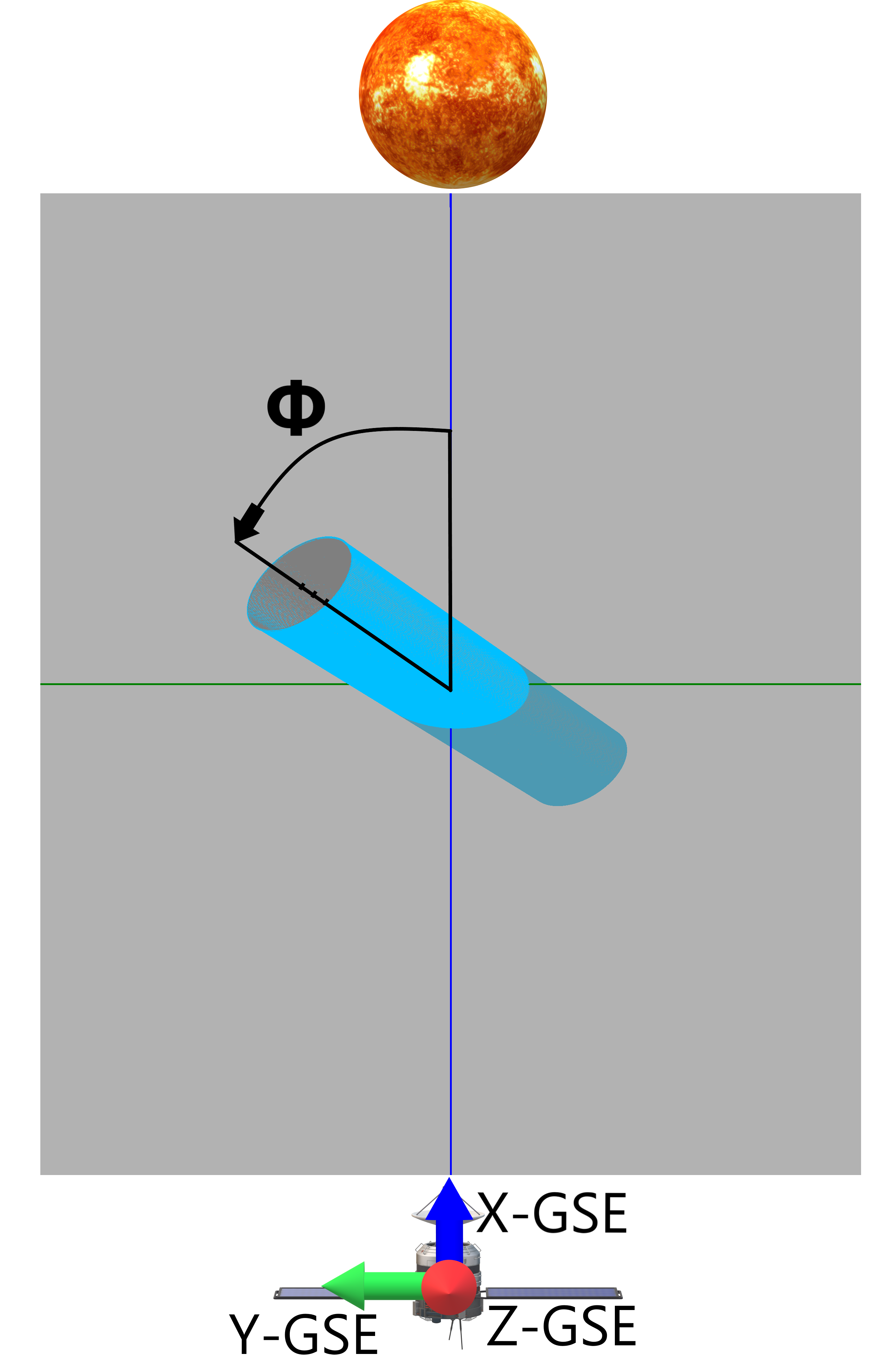}}
\centerline{\textbf{c}\includegraphics[height=0.45\textwidth,clip=]{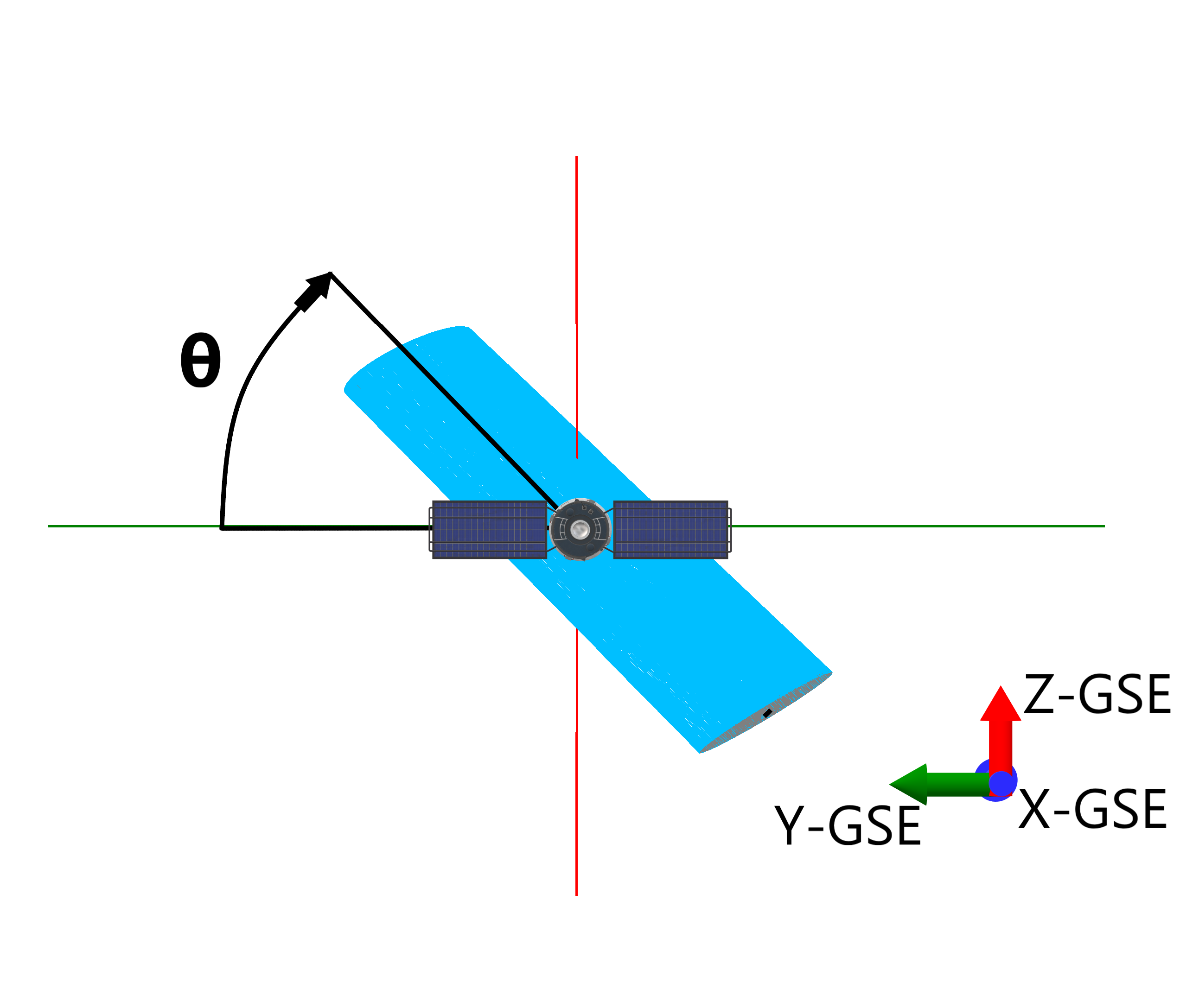}
            \textbf{d}\includegraphics[height=0.45\textwidth,clip=]{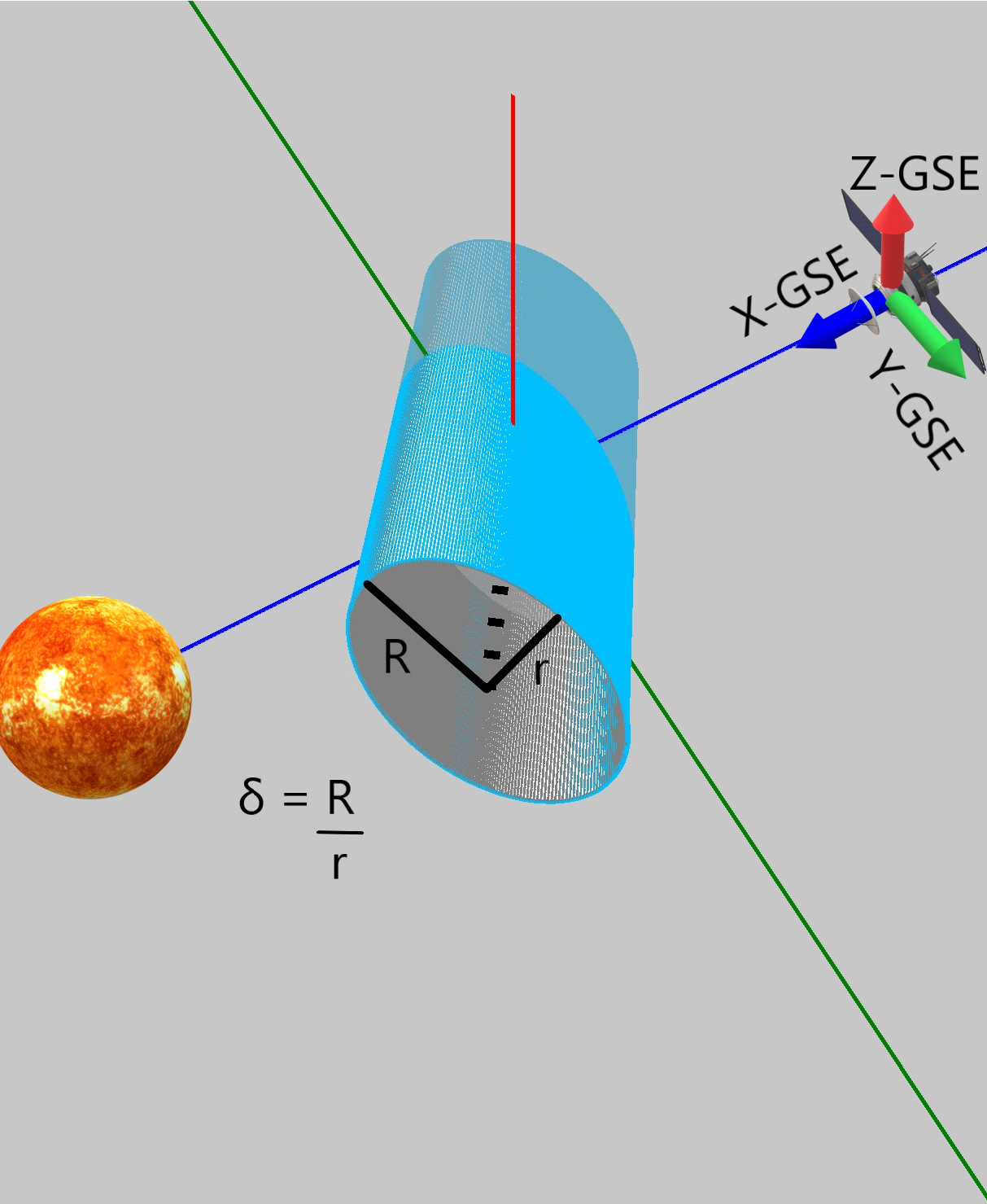}}

          \caption{Flux rope example generated using the Elliptic-cylindrical model using the parameters $\phi$ = 60, $\theta=45$, $Y_{0}=0$, $\xi=40$, $\delta = 0.5$ and $H=+1$ in GSE coordinate system. (a) Overview of the Flux Rope from the point of view out of the ecliptic plane (plane XY) showing the $\xi$ rotation about the central axis. (b) View of the flux rope along the Z-axis. (c) View from the Earth to Sun direction (i.e. the spacecraft point of view). In this case, $Y_0 = 0$ indicates the spacecraft goes through the flux rope central axis. (d) View of the cross-section of the flux rope. }
\label{F-fr-model}
\end{figure}

The FR data set is created using the Elliptic-cylindrical model (EC) \citep{Nieves_E_2018}, consisting of time series of each magnetic field components of a simulated spacecraft trajectory through the modelled flux rope. The EC model has eight input parameters:

\begin{longtable}{l p{10cm}}

    $\bm{B_{y0}}$ & The magnetic field at the center of the flux rope, therefore, the maximum magnetic field. We are holding this parameter constant at 10nT since the magnetic fields are all normalized when converted to hodograms. \\
    \\[-0.5em]
    $\bm{C_{10}}$ & We hold this value constant at 1, which imposes a force free structure. \\
    \\[-0.5em]
    $\bm{H}$ & Chirality of the flux rope. We set this to $\pm1$ to produce cases of both left- and right-handed chirality.\\
    \\[-0.5em]
    $\bm{Y_0}$ & The perpendicular distance from the center of the flux rope to the crossing of the spacecraft. For this proof-of-concept stage we are holding this value as zero AU, so all simulations are crossing at the center of the flux rope. \\
    \\[-0.5em]
    $\bm{\phi}$ & Flux rope latitude orientation angle. It is varied from 5$^{\circ}$ to 355$^{\circ}$ in steps of 10$^{\circ}$. Refer to Figure~\ref{F-fr-model}-b \\
    \\[-0.5em]
    $\bm{\theta}$ & Flux rope longitude orientation angle. It is varied from -85$^{\circ}$ to 85$^{\circ}$ in steps of 10$^{\circ}$. Refer to Figure~\ref{F-fr-model}-c  \\
    \\[-0.5em]
    $\bm{\xi}$ & Flux rope rotation about central axis. It is varied from 0$^{\circ}$ to 180$^{\circ}$ in steps of 10$^{\circ}$. Refer to Figure~\ref{F-fr-model}-d \\
    \\[-0.5em]
    $\bm{\delta}$ & The ratio of the two axes of the cross section of the flux rope cylinder. It is varied from 0.2 to 1 in steps of 0.2; 1 giving a circular cross-section and 0.2 a very elliptical cross section. With $\delta$ set to 1, we have a Circular-Cylindrical model. Refer to Figure~\ref{F-fr-model}-d  \\

\end{longtable}
\addtocounter{table}{-1}

For more details about the parameters, please refer to \cite{Nieves_E_2018}. An interactive tool of flux rope configuration parameters is available at https://www.geogebra.org/m/navfskxj. The permutation of these parameters generates a total of 123,120 different synthetic events to be used for training. \\

Figure~\ref{F-fr-model} contains four panels with different views of a Flux Rope obtained using EC model with parameters $\phi = 60$, $\theta=45$, $Y_{0}=0$, $\xi=40$, $\delta = 0.5$ and $H=+1$ and in the GSE coordinate system. The top left figure is an open field view of the Flux Rope from above the ecliptic plane (plane XY) and gives an overview of the orientation of the flux rope relative to the spacecraft and Sun.  We also observe the angle $ \xi $, which is the rotation of the flux rope about its central axis.  The top right panel is North to South view of the flux rope showing the ecliptic plane and the $\phi$ angle, between the projection of the flux rope axis in the ecliptic plane and the x-axis. The bottom left panel has a view from the Earth to Sun direction, and it shows the YZ plane with the $\theta$ angle, the angle between the projection of the flux rope axis in the YZ plane and the y-axis. This panel can also be used to understand the impact parameters ($Y_{0}$) which is the crossing distance from the flux rope axis. In this case, it is zero. The last panel presents the circular cross-section of the flux rope where it is possible to see the two different radii of the ellipse, R and r, which are used to calculate $\delta$. \\

\begin{center}
      \begin{figure}[ht!]  
   \centerline{\includegraphics[width=0.8\textwidth,clip=]{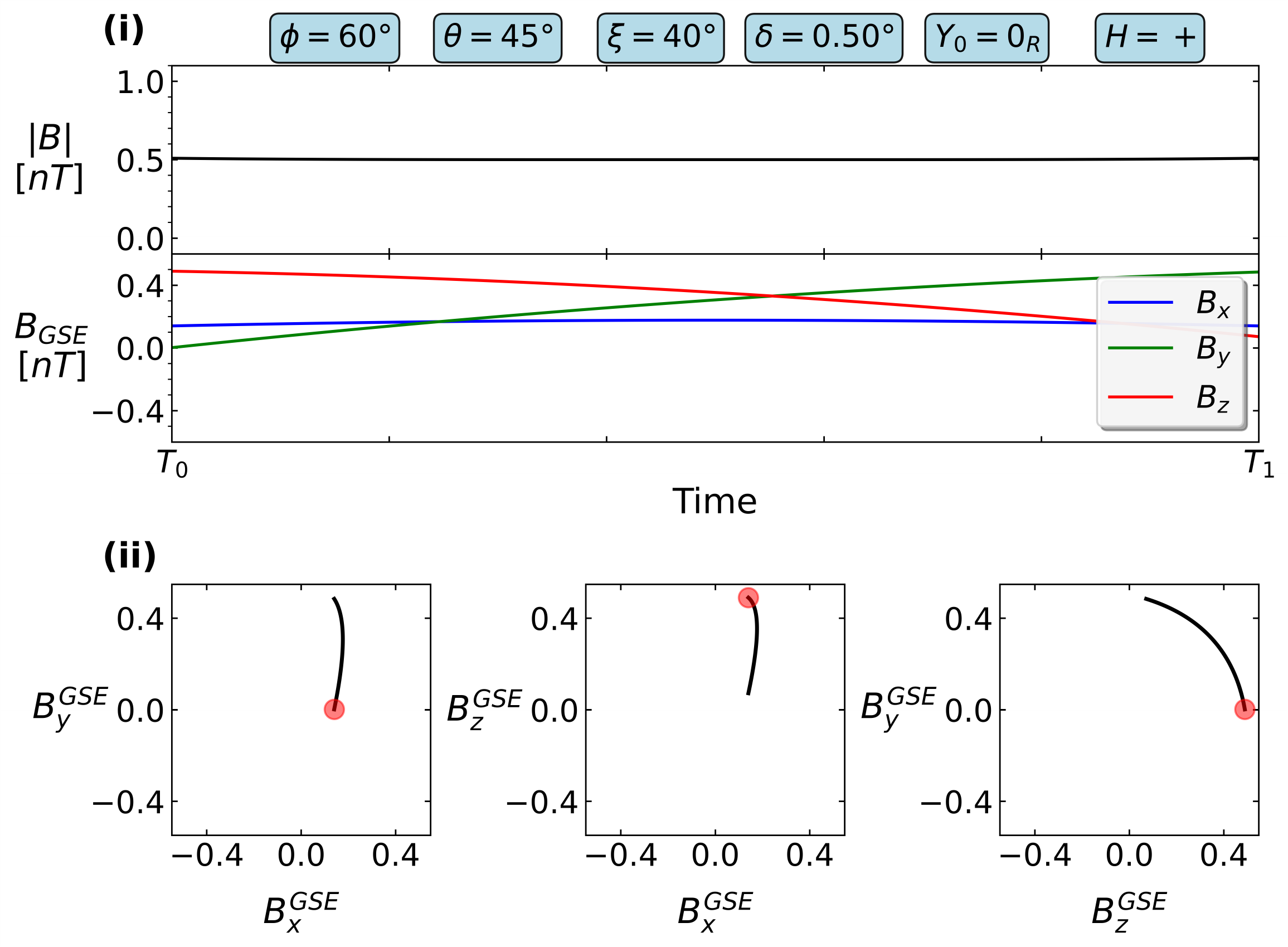}              }
              \caption{A flux rope example generated using the Elliptic-cylindrical model using the parameters $\phi$ = 60, $\theta=45$, $Y_{0}=0$, $\xi=40$, $\delta = 0.5$ and $H=+1$. i) The total magnetic field and the magnetic field components. ii) Three hodograms panels of the magnetic field components. From left, $B_{y}^{GSE}$ \textit{vs} $B_{x}^{GSE}$, $B_{z}^{GSE}$ \textit{vs} $B_{x}^{GSE}$ and $B_{y}^{GSE}$ \textit{vs} $B_{z}^{GSE}$. The red dot represents the starting point of the magnetic field.} 
   \label{F-Example-Synthetic}
   \end{figure}
\end{center}

Figure~\ref{F-Example-Synthetic} exhibits the same example of a synthetic flux rope as shown in  Figure~\ref{F-fr-model}. The top 2 panels include the total magnetic field and the magnetic field components. The $B_y$ component changes from 0 nT to +0.45 nT, while $B_x$ is entirely flat at 0.15 nT, and $B_z$ changes from +0.45nT to 0 nT. We create a flux rope signature image by combining the time series of each magnetic field component in three hodograms by plotting $B_y$ \textit{vs.} $B_x$; $B_z$ \textit{vs.} $B_x$, and $B_y$ \textit{vs.} $B_z$. The three bottom panels are the hodograms of this event, and the red dot is the start point of the magnetic field. Hence, we have three image components for each event that map the 3-dimensional space down to a 2-dimensional signature suitable for input into our DCNN. This process also eliminates the time component, which simplifies the problem to a geometrical identification. We are aware it may be necessary to reconsider time dependence later for future goals like a real-time prediction.\\

The EC model effectively creates positive cases (FRs) with varied combinations of parameters, but creating negative training data (NFRs) holds its own challenges. While the instances of MCs that do not match a flux rope geometry are not well understood, they have been broadly categorized into two groups by \cite{Nieves_2019}, Ejecta (E), and Complex (Cx). Hodograms of the real E cases are more visually distinct from the cases classified as flux ropes than are those from the Cx cases. Since the scope of this research is to demonstrate the analysis of simple events, we have worked with only the E type of non-flux rope event. \\

To create these synthetic ejecta events, we created three time series pulled from a Gaussian distribution. The mean and standard deviation of each time series were selected randomly from uniform distributions in the ranges [-.6,.6] and [.1,.3], respectively. Any points falling outside $\pm 1$ were replaced with the mean.  Each of the time series was treated as one magnetic field component and were plotted as hodograms in precisely the same way as were the positive synthetics. To have a balanced training dataset, we created a total of 123,120 synthetic ejecta events.\\

\begin{center}
      \begin{figure}   
  \centerline{\includegraphics[width=0.8\textwidth,clip=]{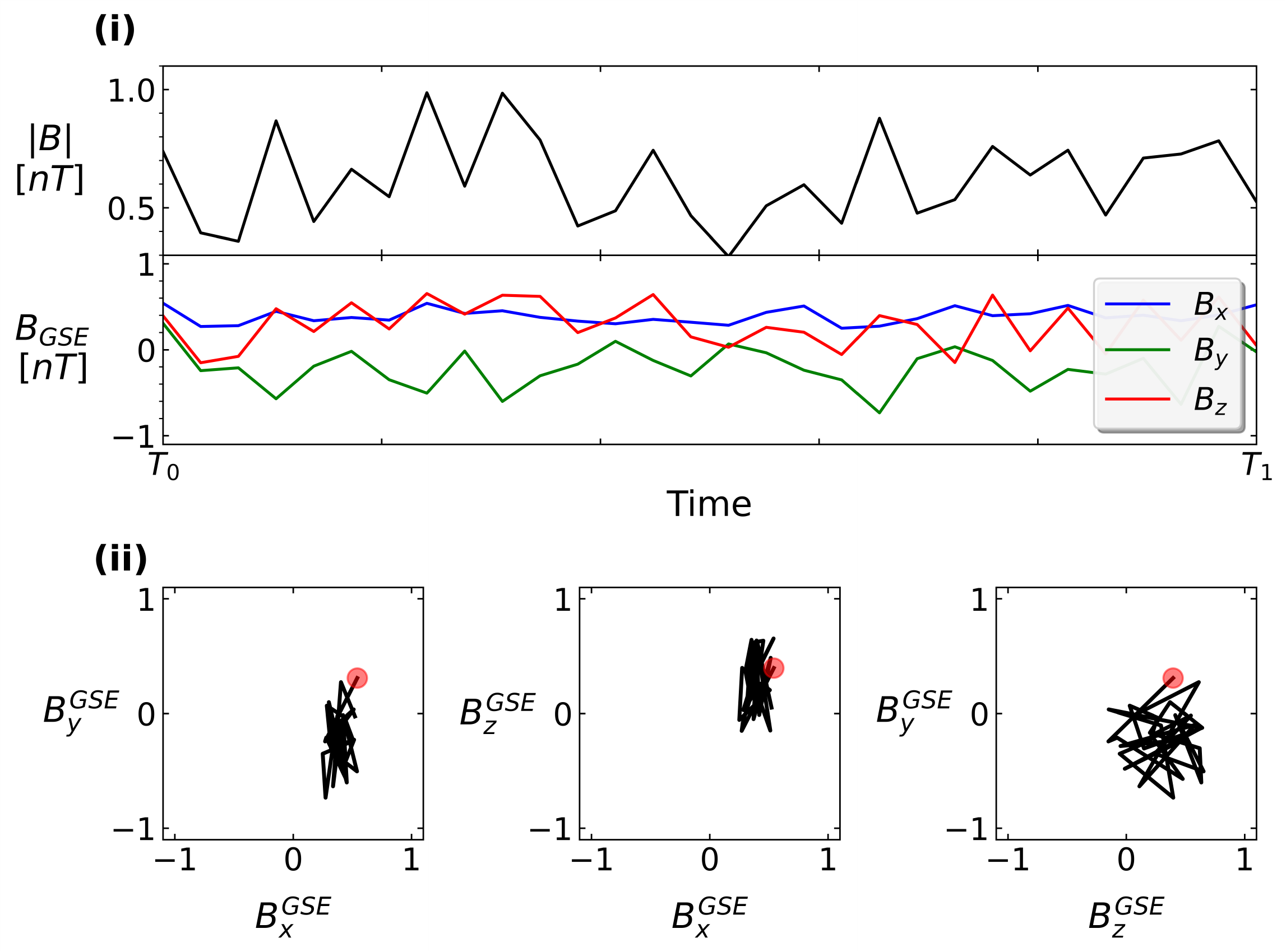}              }
              \caption{
              An ejecta example generated using three time series pulled from a Gaussian distribution. i) The total magnetic field and the magnetic field components. ii) Three hodograms panels of the magnetic field components. From left, $B_{y}^{GSE}$ vs $B_{x}^{GSE}$, $B_{z}^{GSE}$ vs $B_{x}^{GSE}$ and $B_{y}^{GSE}$ vs $B_{z}^{GSE}$. The red dot represents the starting point of the magnetic field.}
  \label{F-sample-ejecta}
  \end{figure}
\end{center}

Figure \ref{F-sample-ejecta} displays an example of synthetic ejecta generated using the Gaussian distribution method. On the top panels are the total magnetic field and magnetic field components, where it is possible to see that there is no clear trend or rotation of any component. On the three bottom panels, there are the hodograms of this event, which also show no evident rotation on any component of the magnetic field.\\

All hodograms are re-scaled and plotted in the same range.  We introduced a small, white margin, and all image files are created at a resolution of 32 x 32 pixels for training and analysis. Two additional sets of training data were created following the same methodology but with two different noise levels added. The set called ``5\% Noise" added values drawn from a Normal distribution with a standard deviation of 0.05 to each data point. Likewise, the ``10\% Noise" set uses a standard deviation of 0.1. \\
 
\subsection{Real Data}
    \label{S-Dataset}
 
This work uses the catalog published in \cite{Nieves_2018}. The classification done in the paper \cite{Nieves_2019} was based on the rotation of magnetic field components of each event. The events which do not show any apparent rotation of the magnetic field components are classified as Ejecta (E). Events with evident rotation are classified as $F^-$, $Fr$, or $F^+$, depending on the span of the rotation. Events with more complex rotations of the magnetic field components, more than 270 degrees or more distinct structures, are classified as Complex (Cx). For our purpose, we hold the \citeauthor{Nieves_2019} classification to its broader level, considering all cases of type $F^-$, $Fr$, and $F^+$ as flux rope (FR) and all cases of E and Cx as non-flux rope (NFR) \\

\begin{figure} [ht]   
   \centerline{
                \hspace*{-0.03\textwidth}
                \includegraphics[width=0.8\textwidth,clip=]{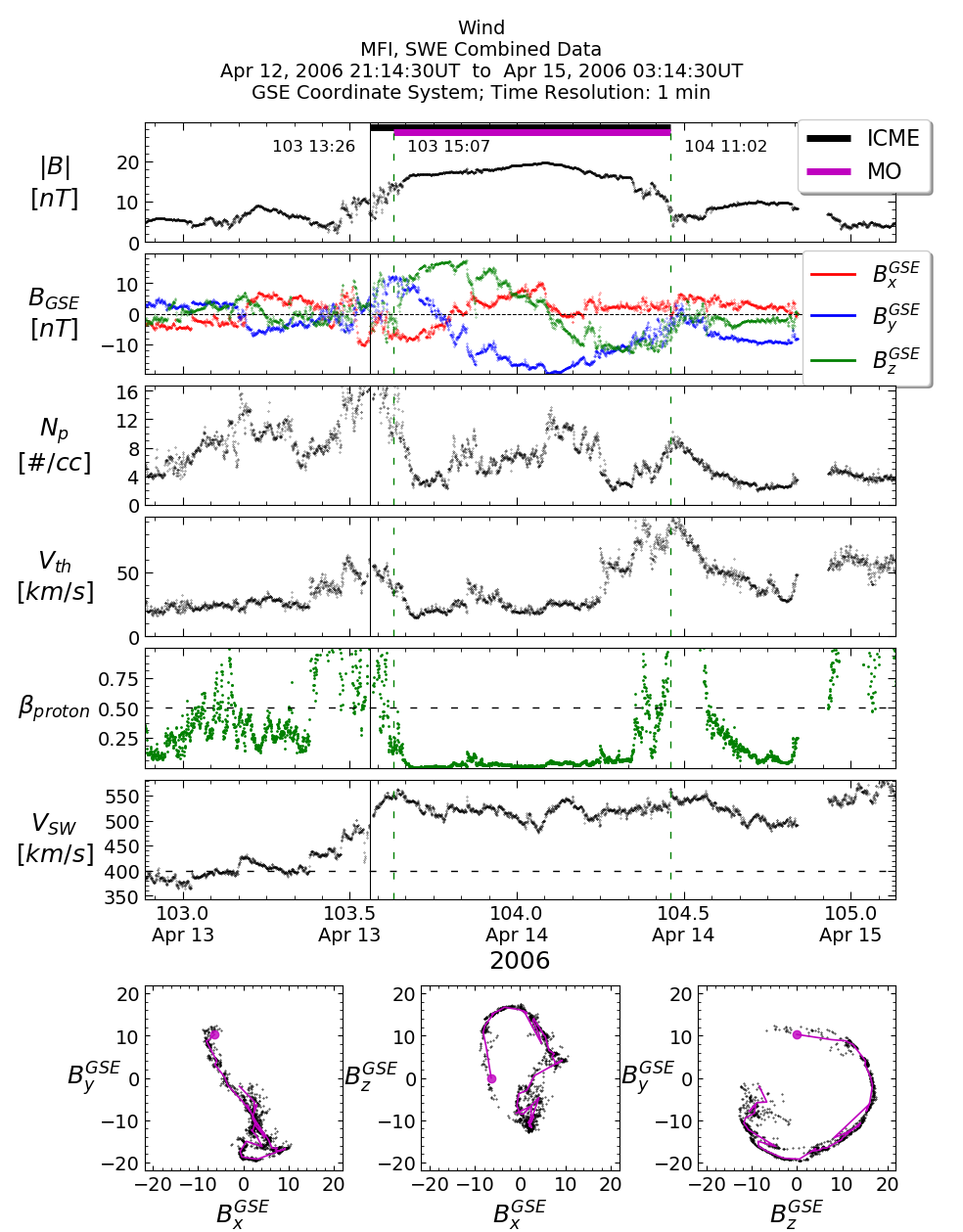}
              }
     \vspace{-0.35\textwidth}   
     \centerline{\Large \bf     
         \hfill}
         
     \vspace{0.31\textwidth}    

             \caption{From top to bottom, Total Magnetic Field [nT], Magnetic Field Components [nT], Proton Density [\#/cc], Thermal Velocity [km/s], $\beta_{proton}$ (ratio of gas pressure and magnetic pressure), Bulk Velocity [km/s] and three hodograms of the magnetic field components.ICME observed on April 13, 2006, classified as ``Fr" Flux Rope.}                    
   \label{F-DatasetExample-1}
   \end{figure}

\begin{figure} [ht]   
   \centerline{
                \hspace*{-0.03\textwidth}
                \includegraphics[width=0.8\textwidth,clip=]{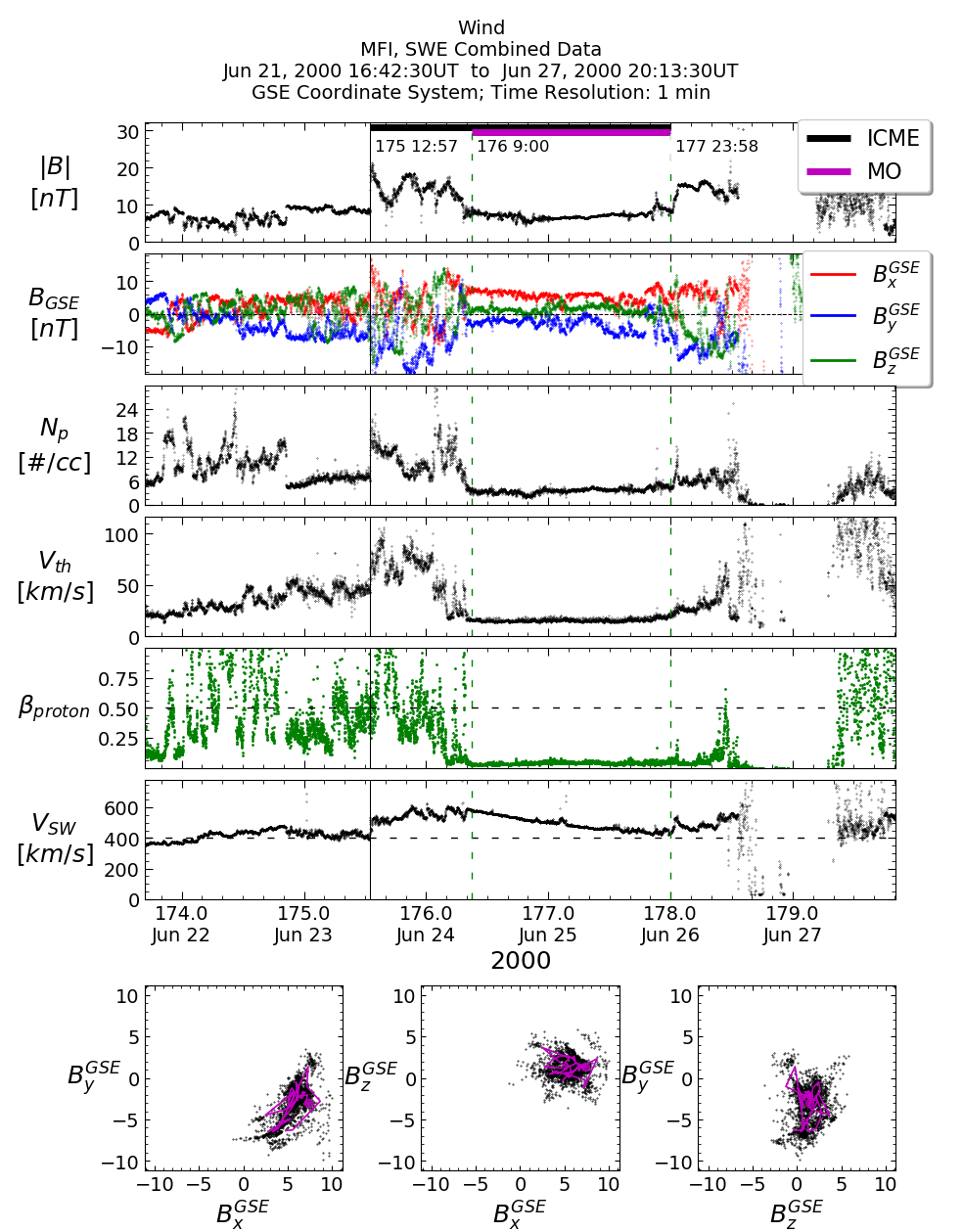}
              }
     \vspace{-0.35\textwidth}   
     \centerline{\Large \bf     
         \hfill}
         
     \vspace{0.31\textwidth}    

             \caption{From top to bottom, Total Magnetic Field [nT], Magnetic Field Components [nT], Proton Density [\#/cc], Thermal Velocity [km/s], $\beta_{proton}$ (ratio of gas pressure and magnetic pressure), Bulk Velocity [km/s] and three hodograms of the magnetic field components.  An ICME From June 23, 2000, classified as ``Ejecta."}
   \label{F-DatasetExample-2}
   \end{figure}

Figures \ref{F-DatasetExample-1} and \ref{F-DatasetExample-2} show two examples of events from the Nieves-Chinchilla catalog. The Figure \ref{F-DatasetExample-1} event is an ICME observed on April 13, 2006 and classified as FR. The classification was based on the smooth and clear rotation of the $B_y$ and $B_z$ components, while the $B_x$ tends to be closer to zero. The event shown in Figure \ref{F-DatasetExample-2} was observed on June 23, 2000 was classified as E. No clear rotation is seen in this event and all three components are approximately flat although it displays a coherent configuration in magnetic field and the other quantities like thermal velocity, proton density, and $\beta_{proton}$ (the ratio of gas pressure and magnetic pressure), which is a signature associated with MOs.\\

The reference catalog has 353 events. Of these, 32 are used during the select evaluation phase, and the remaining 321 are reserved for results analysis in section \ref{S-results}. We selected 32 cases (indicated with ``*" in Table \ref{T-complete-events}) as being most easily differentiated by eye as FRs or NFRs and considered a good test-bed in which to evaluate the performance of a machine-learned based classifier. The FR cases are some of the most smooth and ``nicely" behaved events, while the NFR cases were all of the subset type E. \\


Each magnetic field component is averaged to one hour time intervals unless the result holds fewer than 20 points, in which case we move to a smaller time window for averaging. Because this work is focused on the geometry of magnetic structures and not the magnitude, all events are re-scaled and plotted in the same range in hodogram format. All image files are created at resolution 32 x 32 pixels for neural network training and evaluation.\\

\subsection{DCNN Training Pipeline}
\label{S-Model-Evaluation}
Initially, we set up several similar DCNN-model architectures. Each of these neural networks was trained with 128 noise-free, synthetic events per batch, withholding a randomly selected 30\% of the training data for validation. For this training of network weights, we used the Adam optimizer \citep{Optimizer} with an initial learning rate of 0.001. We found accuracy and loss converging quickly and suspected it was due to the simplicity of this classification problem in simulation space. Thus, we limited training of the network to 50 epochs to avoid overfitting.\\

The most meaningful result we have is not how well we can train the DCNN-model to recognize the differences between the synthetic data but how it performs when classifying the real events. Accordingly, after 50 epochs of training, we evaluated each of the considered DCNN-model architectures with our selected real cases (section \ref{S-Dataset}) and predicted their label. From this, we can score our DCNN-model's overall accuracy in its desired use-case. The best performing epoch and architecture, presented in subsection \ref{S-machine-learning-model} was selected for further development.\\

With the DCNN-model architecture selected, we began DCNN-model refinement using the three synthetic training datasets and secondary evaluation with the selected real data. As before, the DCNN-model has its network weights trained using noise-free synthetic data. After each epoch of training, we validate the DCNN-model against our selected real cases. Keep in mind, this evaluation does not have any feedback to the model weights and it serves only as a parameter for model optimization through selection.\\

We then extend the DCNN-model's training in a step-wise manner by introducing noisy training data. Using the epoch giving the best performance on the real events as our trained source model, we create a new copy of the ML model and initialize its convolutional layers with the learned network weights from the source model, while randomizing the weights on the fully connected layers, as described in \cite{barshan2015stage}. We then trained this new model for 50 epochs with the 5\% Noise training set. In a like manner, we then extended the training from the best performing epoch of this 5\% model, this time training with the 10\% Noise dataset. In this way, the secondary and tertiary stages can build on the spatial relationships learned in earlier stages while allowing for new classification criteria better-aligned with noisy input data. Each stage of training creates a separate DCNN-model that can be evaluated independently against real-world data.\\

\subsection{DCNN-model Evaluation}
    \label{S-real-validation}

Table \ref{T-confusion-matrix} displays metrics extracted from the classification results of evaluation using the real data subset. Cases where the reference and original classification agree on being positive (FR) or negative (NFR) are True Positives (TP) and True Negatives (TN), respectively. If the classification is positive (FR)/negative (NFR) and the ground truth is negative (NFR)/ positive (FR), we have false positive (FP)/false negative (FN). The Accuracy is the ratio of the True (TP+TN) cases to the total number of cases. In addition to TP, FN, TN, FP and Accuracy, the table includes the calculated quantities ``Precision", ``Recall" and ``F1 Score", standard metrics in ML, defined in equations \ref{E-precision}, \ref{E-recall} and \ref{E-fone} respectively in the appendix \ref{A-Metrics}. \\

\begin{table}[ht!]
\caption{Metrics for the classifications made during the training phase using the 32 cases evaluated. It presents these metrics for the three different levels of noise (noise-free, 5\%, and 10\% noise).}
 \begin{tabular}{@{\extracolsep{\fill}}lccc}\hline \hline 
 Quantities & No Noise   & 5\% Noise  & 10\% Noise \\
 \hline
True Positive  & 16 (89\%)  & 18 (100\%)  & 16 (89\%) \\
False Positive &  2 (14\%)  &   5 (36\%)  &  7 (50\%)\\
True Negative  & 12 (86\%)  &   9 (64\%)  &  7 (50\%)\\
False Negative &  2 (11\%)  &    0 (0\%)  &  2 (11\%)\\
Overall Accuracy& 88\%      &  84\%       &  72\% \\
Precision & 0.89 & 0.78 & 0.70 \\
Recall & 0.89 & 1.00 & 0.89\\
F1 Score & 0.89 & 0.88 & 0.78\\

\hline  \hline
\end{tabular}
\label{T-confusion-matrix}
\end{table}

A more detailed classification of the 32 events used in the validation can be found in Table \ref{T-complete-events} (indicated with ``*"), Appendix \ref{A-complete-table} and it has the necessary information to compare the classification done in the reference catalog and the classification done for the DCNN-model with different amounts of noise.\\

Analyzing the results of Table \ref{T-confusion-matrix}, we can see that the DCNN-model works well across all three levels of noise, with high F1 Score, Recall and Precision of 0.89 for the no noise model and 88\% accuracy. These numbers drop to 0.88 for F1 Score and 0.78 for Precision when adding 5\% noise, showing a worse classification of the NFR cases, but a better FR classification with the increase of the Recall to 1. The general accuracy decreased a little to 84\% with the 5\% noise DCNN-model. This tendency remains in the 10\% noise results, with a Precision of 0.7 and F1 Score 0.78, demonstrating an even worse classification of the NFR cases and an overall accuracy of 72\%. The results display a good performance because the model and training data were optimized for this set of thirty-two events, although none of the real events were used in the actual training of the DCNN-model weights. These numbers represent the capability of our model to identify real flux ropes although being only trained with synthetic data.\\

\begin{center}
    \begin{figure} [ht]   
   \centerline{\includegraphics[width=0.8\textwidth,clip=]{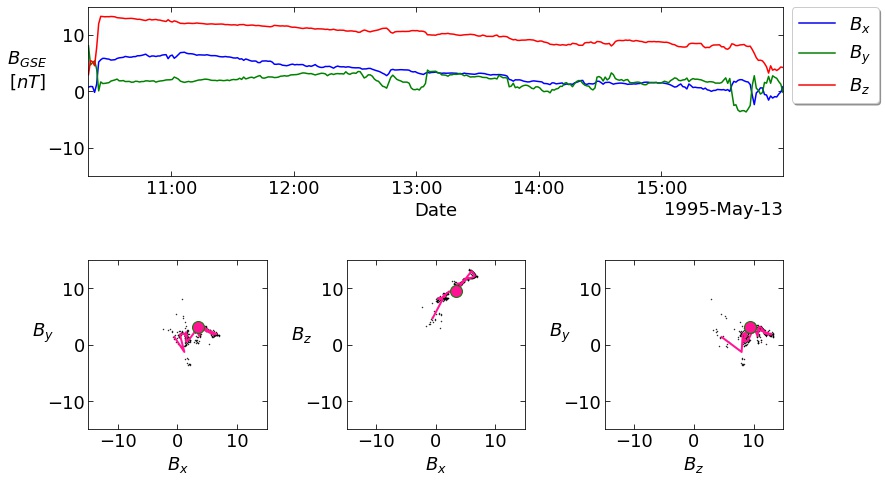}}
    \caption{ICME of May $13^{th}$, 1995. The top panel shows the magnetic field components. The three bottom panels are the hodograms for this event, composed by the real data (dotted) and the smoothed real data (pink line).
 }
   \label{F-event05131995}
   \end{figure}
\end{center}

\begin{center}
      \begin{figure} [ht]   
      \centerline{\includegraphics[width=0.8\textwidth,clip=]{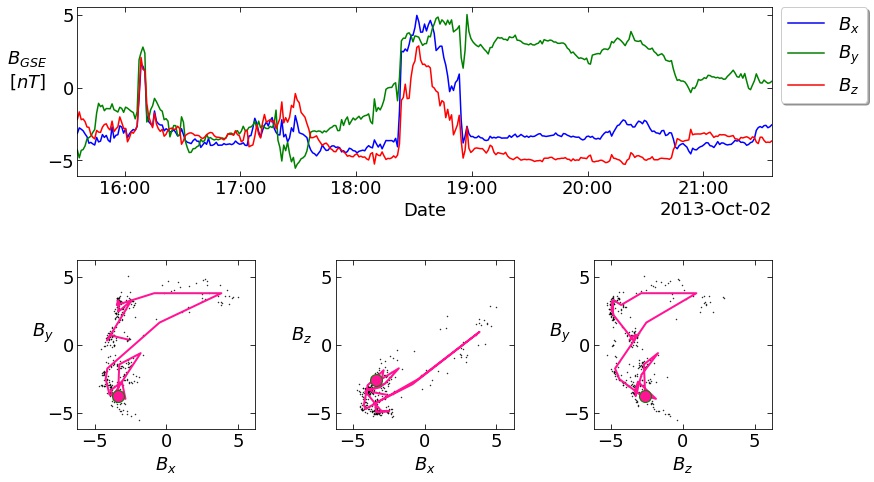}              }
              \caption{ICME of October $2^{nd}$, 2013. The top panel shows the magnetic field components. The three bottom panels are the hodograms for this event, composed by the real data (dotted) and the smoothed real data (pink line).}
  \label{F-event10022013}
  \end{figure}
\end{center}

\begin{center}
      \begin{figure} [ht]   
   \centerline{\includegraphics[width=0.8\textwidth,clip=]{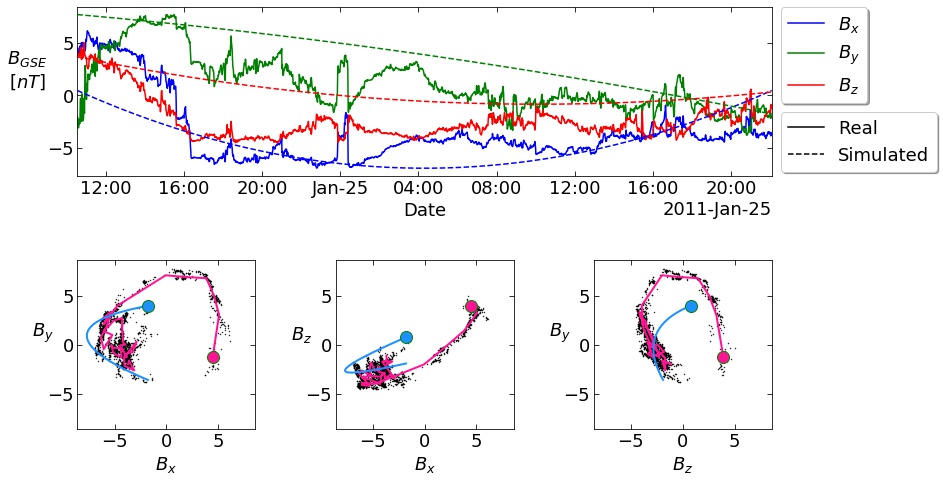}
              }
              \caption{ICME of January $24^{th}$, 2011. The top panel shows the magnetic field components. The solid line is the observed data, and the dashed line is the fitting done using the Circular-cylindrical model \citep{Nieves_C_2016}. The three bottom panels are the hodograms for this event, composed by the real data (dotted), the smoothed real data (pink line), and the fitting done using the CC model (blue line).}
   \label{F-event20110124}
   \end{figure}
\end{center}

\begin{center}
      \begin{figure} [ht]   
   \centerline{\includegraphics[width=0.8\textwidth,clip=]{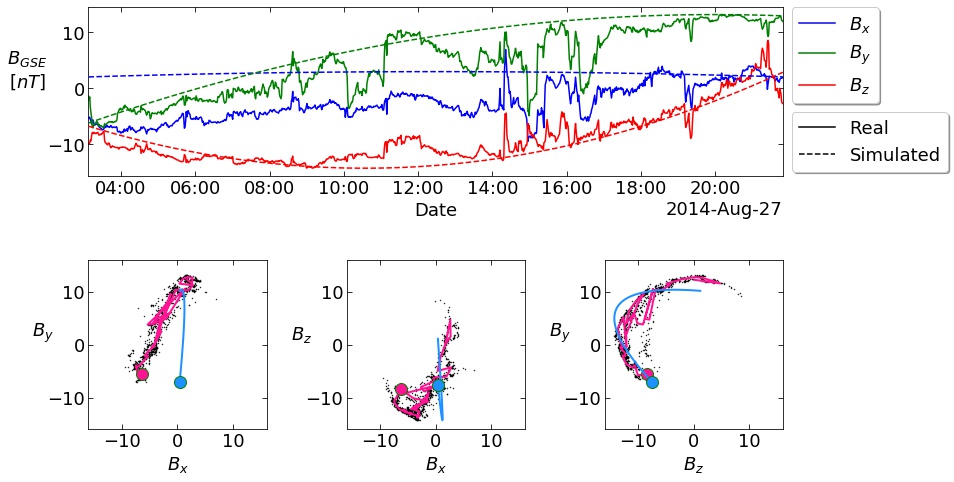}
              }
              \caption{ICME of August $26^{th}$, 2014. The top panel shows the magnetic field components. The solid line is the observed data, and the dashed line is the fitting done using the Circular-cylindrical model \citep{Nieves_C_2016}. The three bottom panels are the hodograms for this event, composed of the real data (dotted), the smoothed real data (pink line), and the fitting done using the CC model (blue line).}
   \label{F-event20140826}
   \end{figure}
\end{center}

Evaluating the noise-free classification results, we found four disagreements out of thirty two cases classified. Figures \ref{F-event05131995} through \ref{F-event20140826} display the four disagreement events which are May $13^{th}$ 1995 (FP), October $2^{nd}$ 2013 (FP), January $24^{th}$ 2011 (FN) and August $26^{th}$ 2014 (FN) respectively.  The top panel has the magnetic field components time series. The three bottom panels are the hodograms for each event, composed of the real data (dotted) and the smoothed real data (pink line). Here we make a detailed evaluation of the classification done in these events.\\

The event of May $13^{th}$, 1995 (Figure \ref{F-event05131995}) was originally classified as ``Ejecta" by \cite{Nieves_2019} while the DCNN-model classified it as ``Flux Rope." This event has a relatively short duration, and by visual inspection of the hodograms, it is clear that it is, to some extent, well behaved. All the components have linear behavior, in addition to the monotonous decay of the $B_z$ and $B_y$ components. Because of the smooth but short rotation in the magnetic field, mainly the $B_z$ component, a case could be made that the reference classification in this instance could be reconsidered as type $F^-$. Alternatively, the negative synthetic data created to train the model is based on Gaussian distributed random numbers, and it may not represent all the ejectas well, as in this case. Implementing more complexity in the synthetic ejecta may address this discrepancy.\\

The event of October 2013 (Figure \ref{F-event10022013}) is the second FP case, where the DCNN-model classified it as an FR, disagreeing with the catalog that classified it as ``Ejecta." It is possible to observe a substantial change in all three components, but mainly  $B_z$ and  $B_x$, at about the halfway point of the event. While there is a small rotation in the $B_y$ component, the hodogram signature is again clearly not well-fit by our simulated negative training data and could also benefit from implementing a more complex negative data generator. 

The event of January $24^{th}$, 2011 (Figure \ref{F-event20110124}) is an FN case, classified as ``Not Flux Rope", with a reference catalog classification of ``Flux Rope." It is possible to see the rotation of $B_x$ and $B_z$ components. Doing a visual inspection, the catalog classification of flux rope seems reasonable with the long, smooth rotation in the magnetic field. This case, however, has a catalog sub-classification of $F^+$, defined as structures that have a rotation of more than 180 degrees in at least two components of the magnetic field. We can explain an \textit{in situ} signature like this if the spacecraft is crossing in one of the flanks of the CME, assuming a croissant shape as described at \cite{Nieves_C_2016}. This kind of event is not modeled with the Elliptic-cylindrical model; therefore, the synthetic data used in this experiment does not produce any event with such significant rotation. It is clear that the flux rope model does not fit the data well, showing a limitation of the generated data used for training. It makes sense that the lack of a global model that represents all possible events reduces the classification model's accuracy. Incorporating a flux rope model that assumes a croissant shape is a desired future step in training this classification tool.\\

Even though the model used to generate the training data does not reproduce $F^+$, our analysis of this event suggests another possible solution or fix to this discrepancy between classification and label. After a careful inspection, we do not agree with the boundaries applied to this event and have concluded that this might be better labeled as a Complex event. When looking at the time series from this event, it is possible to see a discontinuity just at the start of January $25^{th}$. Therefore this event could be split into two flux rope events and considered separately by the flux rope model fitting and machine learning classifier. \\

The last FN event is August $26^{th}$, 2014, seen in Figure \ref{F-event20140826}. The reference catalog labels this as FR, neither under- nor over-rotated. By eye, this seems like it should have been easy to classify. A possible explanation is that the nature of the noise in the data may have contributed to the misclassification, having a large change in noise distribution throughout the flux rope crossing. The neural network classification model still needs continued tuning and augmented training data to increase its precision and make it a more generalized model.\\

\section{Results \& Discussion}
    \label{S-results} 
This section introduces the results of the classification made by the DCNN-model of the remaining 321 events, with and without Cx events, from the reference catalog \cite{Nieves_2019}, and we analyze the metrics obtained on these classifications.\\

\begin{figure}[!ht]
\centerline{\includegraphics[height=0.35\textwidth,clip=]{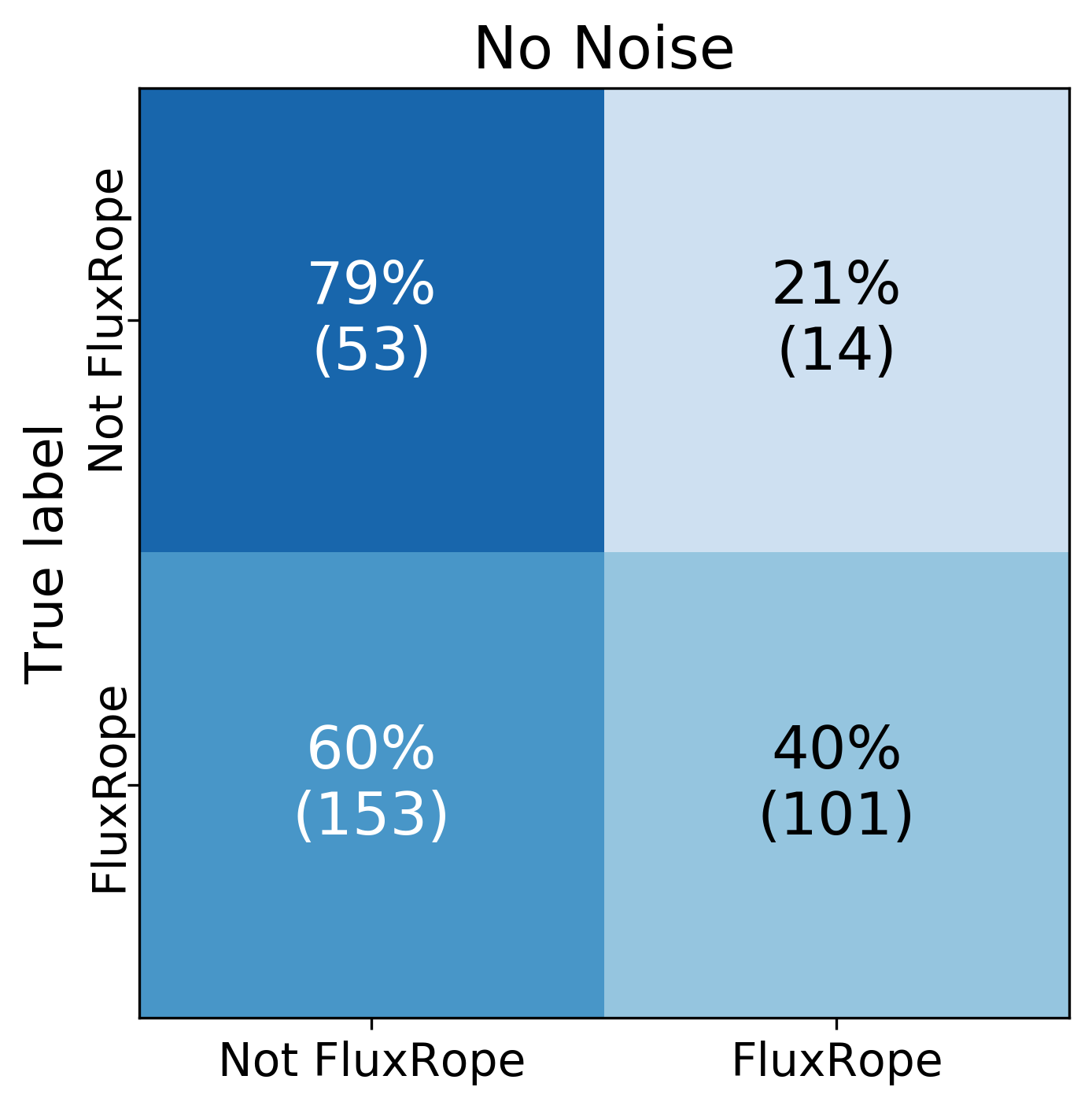}
            \includegraphics[height=0.35\textwidth,clip=]{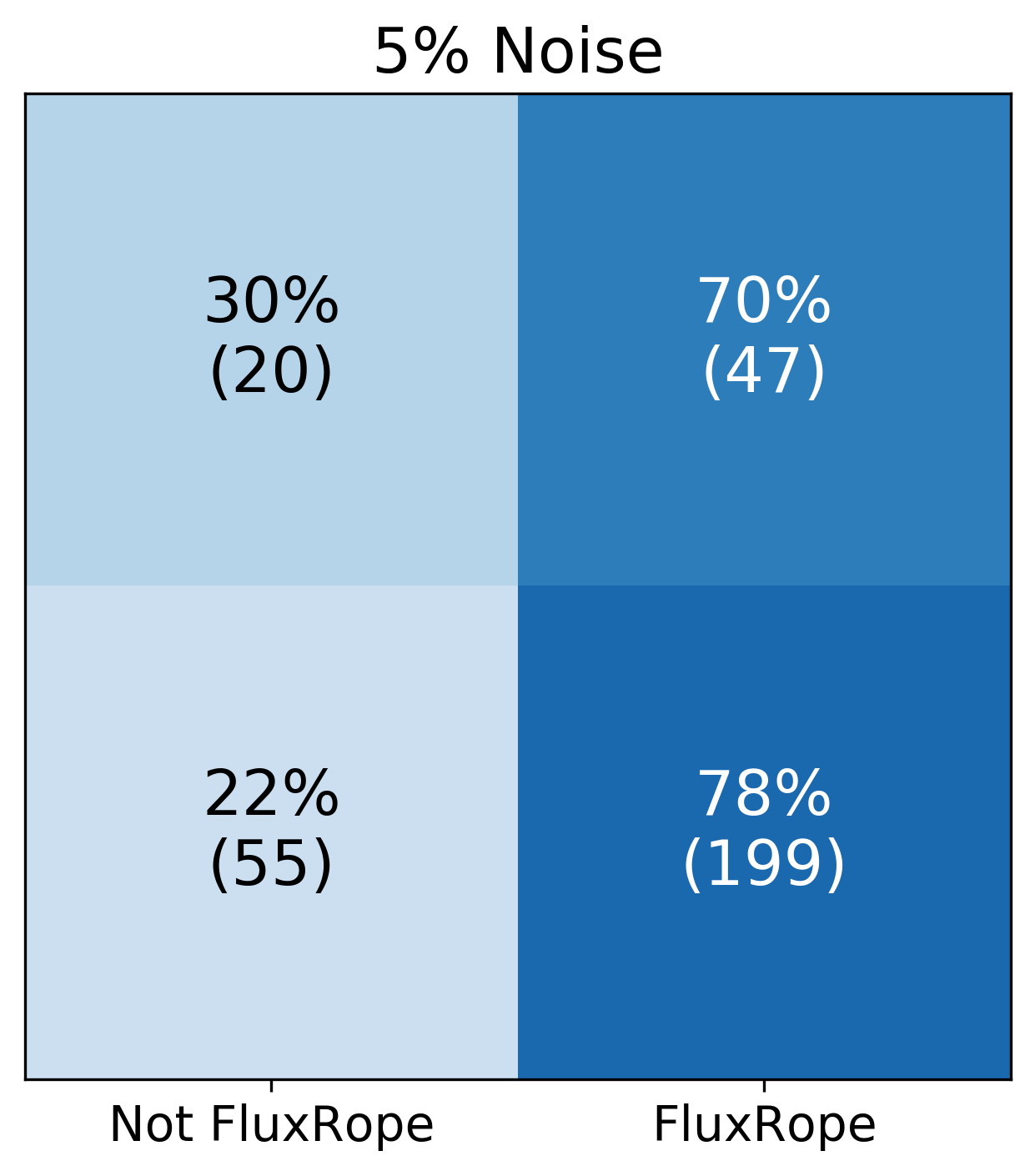}
            \includegraphics[height=0.35\textwidth,clip=]{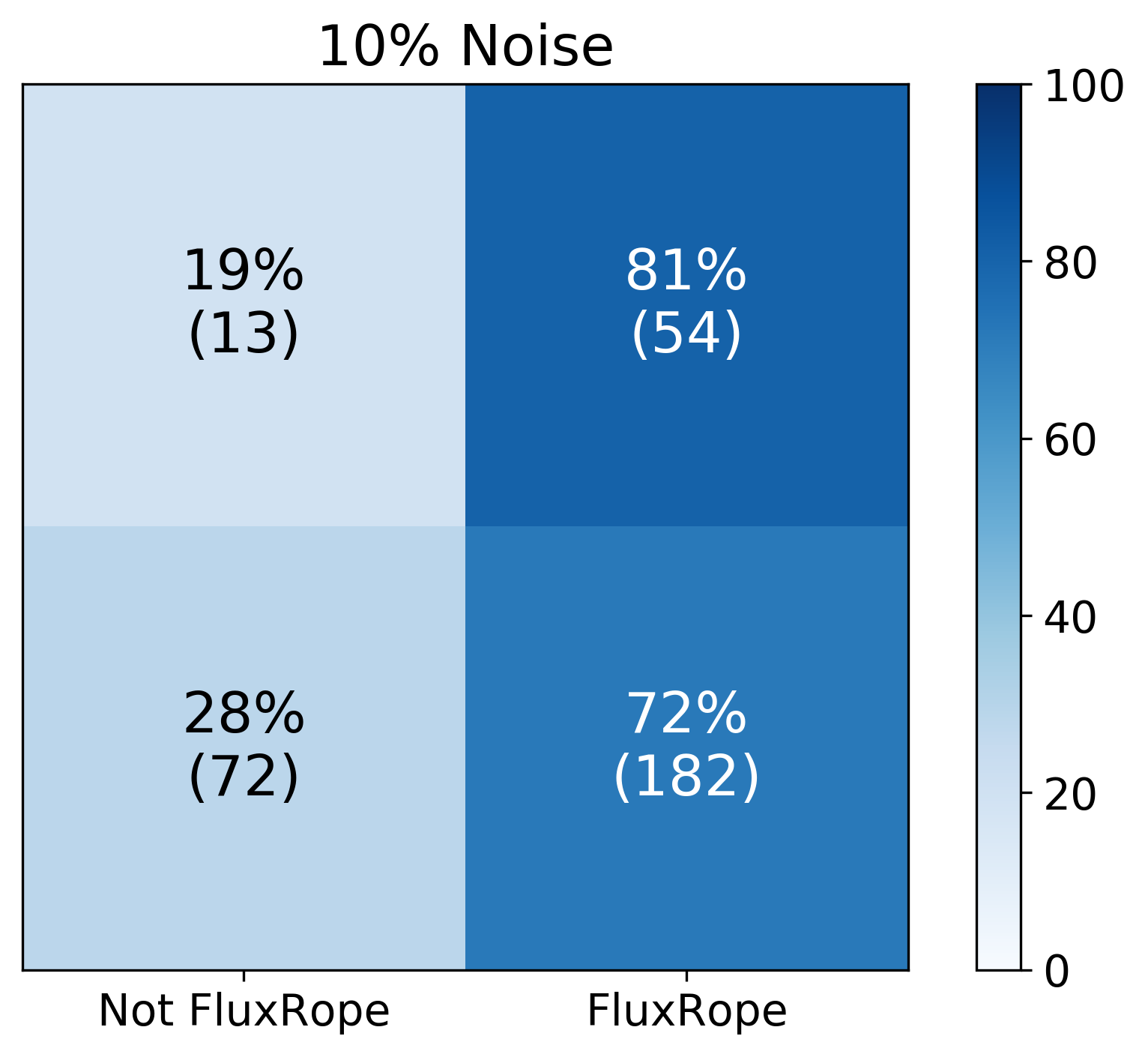}}
\centerline{\includegraphics[height=0.35\textwidth,clip=]{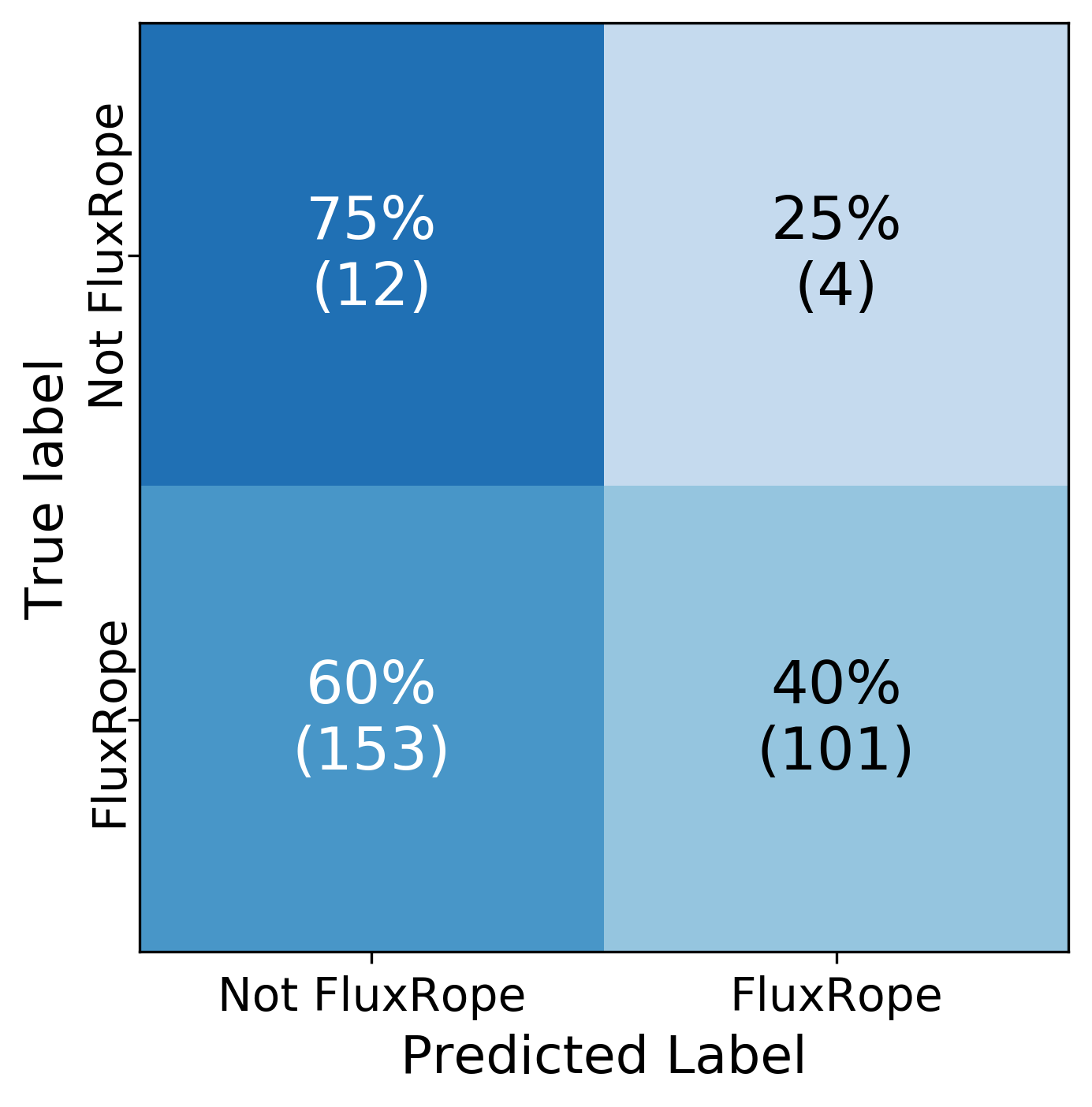}
            \includegraphics[height=0.35\textwidth,clip=]{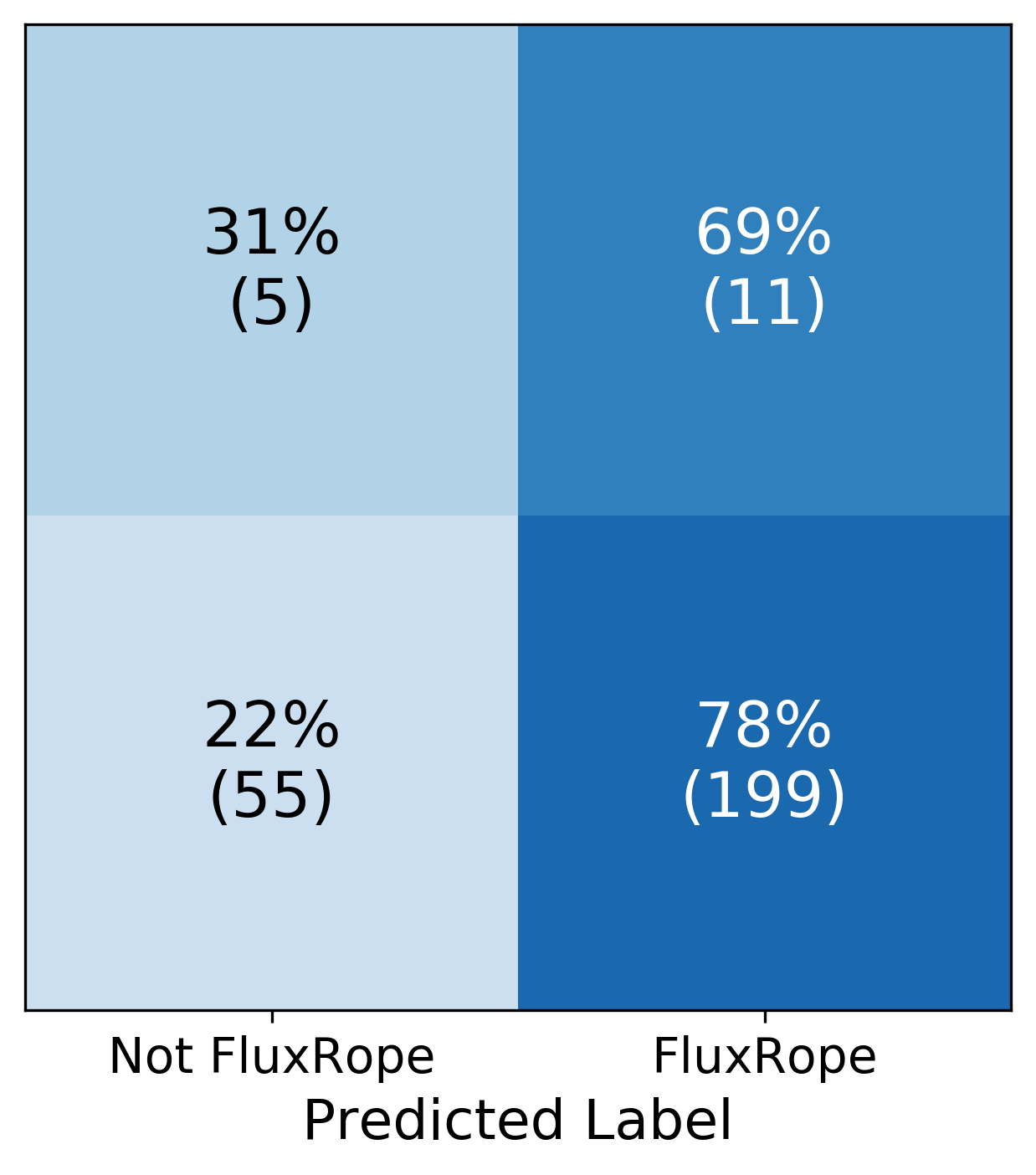}
            \includegraphics[height=0.35\textwidth,clip=]{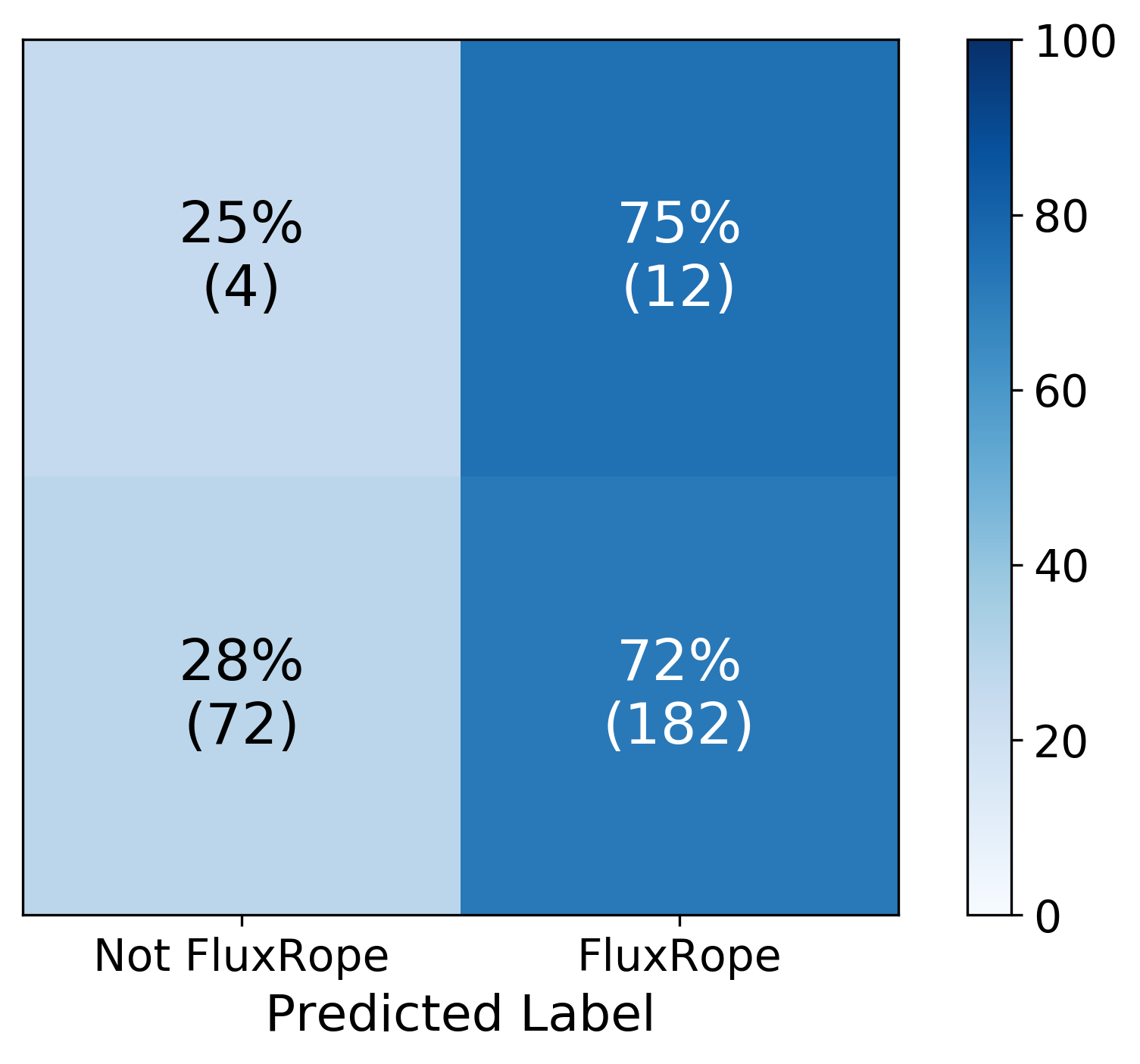}}
          \caption{Six Confusion Matrices (CMs). Each is composed of NxN entries, comparing the true labels and predicted labels of the classified objects. The top row CMs are results when evaluated at the 321 events, including the complex (Cx) structures events. The bottom row is the results when evaluated only at the 270 events that are not cataloged as Cx type. Each CM is for each model trained with different amounts of noise, from left to right CM for the noise-free model, for the 5\% noise model, and the 10\% noise model.}
\label{F-confusion-matrix}
\end{figure}

In Figure \ref{F-confusion-matrix} there are six Confusion Matrices (CM), a.k.a ``error matrix" \citep{Stehman_1997}, used to better visualize the classifier performance. Each is composed of NxN entries, comparing the true labels and predicted labels of the classified objects. In our case, we have only two classes, FRs and NFRs. Each CM represents the DCNN-model trained with different amounts of noise, so from left to right, the first column of CMs are for the noise-free model, second for the 5\% noise model, and the last one for the 10\% noise model. The top row CMs are results when evaluated at the 321 events, including the complex (Cx) structures. The bottom row are the results when the Cx are not included, which amounted to 270 events. The y-axis is the ``True Label," and the x-axis is the ``Predicted Label," and each cell of the CM represents a different quantity. We have the true negatives (TN) in the top-left cell, true positives (TP) in the bottom-right cell, the false-positives (FP) in the top-right cell and the false-negative (FN) case in the bottom left cell.\\

Table \ref{T-confusion-matrix-extended} recreates the information included in Table \ref{T-confusion-matrix} but here based on the 321 events including Cx and 270 events not including Cx. It presents extracted quantities from the confusion matrices from all six ML models evaluated and some complementary metrics to understand the classifications. According to the accuracy in Table \ref{T-confusion-matrix-extended}, the results from the noise-free synthetic data indicate the DCNN-model can predict 79\% of Non-flux ropes correctly when we include Cx events and has a precision of 0.88 but is only correct in 40\% of labeled flux rope cases, resulting in low Recall and F1 Score. These numbers get slightly similar when not including the Cx structures. The model predicts 75\% of the Non-flux ropes and has a precision of 0.96, resulting in a low F1 Score. \\

\begin{table}[ht]
\centering
\caption{Table includes the metrics for the classifications made during the training phase using the 321 cases of the reference catalog with three different levels of noise: noise-free, 5\%, and 10\% noise. It includes TP, FP, TN, FN, accuracy, precision, recall, and F1 Score.}
 \begin{tabular}{@{\extracolsep{\fill}}ccccccc}\hline \hline 
  &  \multicolumn{3}{c}{{\centering Including Cx}} & \multicolumn{3}{c}{{\centering Not including Cx}}  \\
  \cmidrule(lr){2-4}\cmidrule(lr){5-7}
   & No Noise & 5\% Noise & 10\% Noise & No Noise & 5\% Noise & 10\% Noise \\
 \hline
TP & 101 (40\%)  & 199 (78\%)  &  182 (72\%) & 101 (40\%)  & 199 (78\%)  &  182 (72\%)\\
FP &  14 (21\%)  &  47 (70\%)  &   54 (81\%) & 4 (25\%)  & 11 (69\%)  &  12 (75\%)\\
TN  &  53 (79\%)  &  20 (30\%)  &   13 (19\%) & 12 (75\%)  & 5 (31\%)  &  4 (25\%)\\
FN & 153 (60\%)  &  55 (22\%)  &   72 (28\%) & 153 (60\%)  & 55 (22\%)  &  72 (28\%)\\
Accuracy & 48\%       &  68\%    &  61\%  & 42\%       &  76\%    &  69\% \\
Precision & 0.88 & 0.81 & 0.77  & 0.96 &  0.95    &  0.94\\
Recall & 0.40 & 0.78 & 0.72 & 0.40       &  0.78    &  0.72 \\
F1 Score & 0.55 & 0.80 & 0.74 & 0.56       &  0.86    &  0.81 \\
\hline  \hline
\end{tabular}
\label{T-confusion-matrix-extended}
\end{table}

With the addition of 5\% noise, the statistic flips, with 78\% agreement of the labeled flux rope cases but only 30\% of TN cases when Cx are included. When Cx cases are not included, we still have 78\% of TP and a slight increase of TN to 31\%. In both cases, we have a high Recall, Precision, and F1 Score, which is better when no Cx structure is used, which is expected since we did not train the model with synthetic complex cases. These improved Recalls with the addition of noisy training data mirrors what we saw in the subset evaluation. In addition to Recall, the other metrics also improve here, suggesting that classification can be improved by including some noise in the training sets. \\

We observed a drop in the performance with the 10\% noise components in the 321 events set, although not as drastically as compared to the evaluation subset. The Precision dropped 0.04 to 0.77, the Recall 0.04 to 0.72, and F1 Score 0.06 to 0.74. The same happens when tested without the Cx cases; Precision dropped 0.01 to 0.94, Recall 0.06 to 0.72, and F1 Score 0.05 to 0.81. The size of the images used in this work are 32x32 pixels, and this may not be enough resolution to explore all the spatial features created when the 10\% noise is applied. Increasing the resolution of the images may allow a better classification with considerably more noise.\\

In both noise cases, it is possible to observe that the DCNN-model is biased towards Flux Rope, as opposed to the no noise DCNN-model which seems to be biased towards classifying as NFR. This explains the decrease in the TN numbers while noise is added. It is clear that when adding noise, the model starts to classify E and Cx as FR since the simulated flux ropes at this noise level have a non-trivial amount of fluctuation; the hodograms start to resemble ejecta and complex cases. This is a known aspect of the project, and more in-depth investigation of the type of noise and its quantity will help to develop better synthetic data for training. A more physical-based noise will be explored for further development of the DCNN-model, to include implementing fluctuations caused by turbulence, waves, or other such physical processes. \\

\begin{figure}[ht]
\centerline{\textbf{a}\includegraphics[height=0.35\textwidth,clip=]{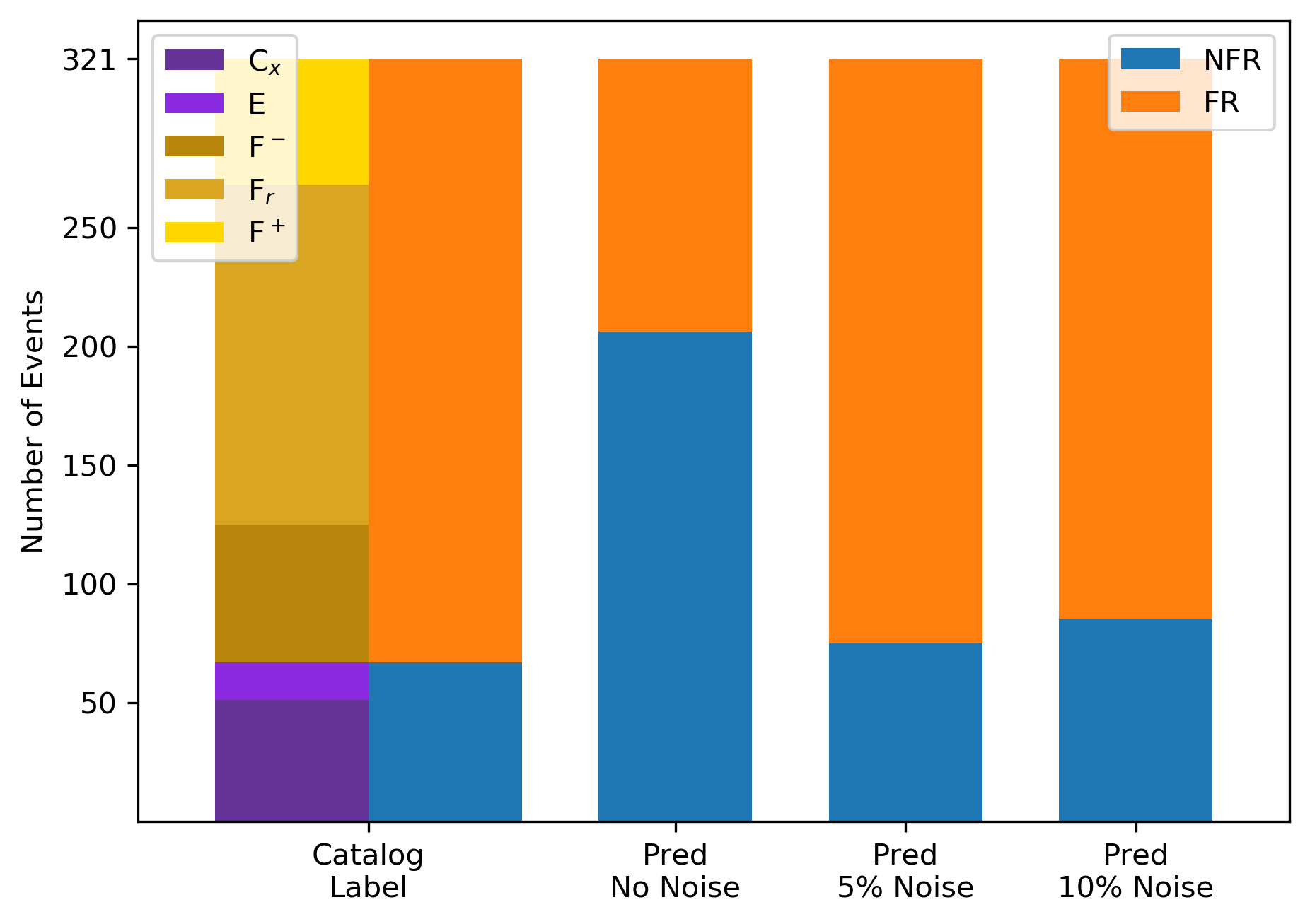}
            \textbf{b}\includegraphics[height=0.35\textwidth,clip=]{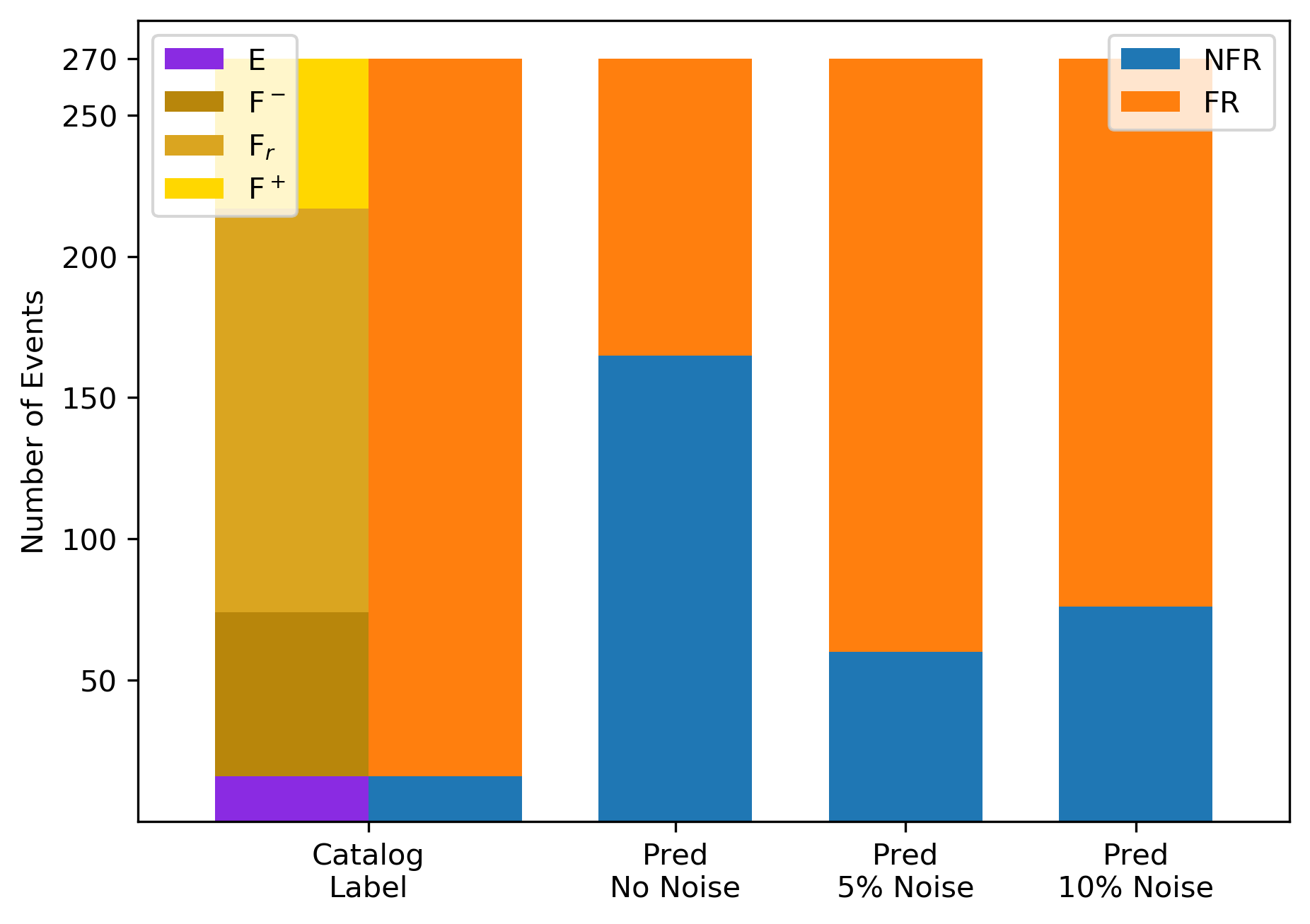}}

          \caption{Two stacked bar plots with the numbers of events in each class from the reference catalog and the predictions made by all three models. (a) Stacked bar plot for the classification, including Cx events. The proportion of NFR in the Catalog label is 67/321 (21\%) and the proportion of FR and NFR is 206/321(64\%) for no noise, 75/321 (23\%) for 5\% noise, and 85/321 (26\%) for 10\% noise. (b) Stacked bar plot for the classification excluding Cx events. The proportion of NFR is 16/270 (6\%) in the Catalog label and 167/270 (61\%) for no noise, 60/270 (21\%) with 5\% noise, and 76/270 (28\%) at 10\% noise.}
\label{F-stacked-bar}
\end{figure}

Figure~\ref{F-stacked-bar} shows two stacked bar plots with the numbers of events in each class from the tested reference catalog and the predictions made by all three models. Panel a is for the classification including Cx events, while panel b is for the classification excluding Cx events. In the 321 events from the catalog used in evalution,  the proportion of NFR is 67/321 (21\%), and it shrinks to 16/270 (6\%) when we remove Cx. The small number of E cases left in the test set was the main reason to include Cx when in our primary classification results; otherwise, the data unbalance is enormous. When we use the Cx cases, the proportion of FR and NFR predicted from the noise trained DCNN-models is close to the reference catalog, 75/321 (23\%), and 85/321 (26\%) respectively. Even though the training data and the validation dataset have a 50\% balance, the classification results still reproduce the reference catalog class ratio. In contrast, the no noise DCNN-model has a very different ratio of NFR, 206/321(64\%), much closer to the ratio of the training data. The ratio of the predictions is approximately the same when we remove the Cx cases, 167/270 (61\%) for no noise, 60/270 (21\%) with 5\% noise, and 76/270 (28\%) at 10\% noise. We know the unbalance in the data is significant and we will add new events from different catalogs that will help with the consequences of the unbalanced data.\\

For a detailed classification of all catalog events, refer to Table \ref{T-complete-events} in Appendix \ref{A-complete-table}. It contains the results for the classification in all 353 cases, with the simple validation subset cases marked with *, and has the necessary information to compare the classification in the reference catalog to the classification in the DCNN-model with different noise levels.\\

The results in Appendix \ref{A-complete-table} demonstrate that the DCNN-model is catching some critical features of flux rope hodograms. It neither classified all the events with a single class nor classified the events randomly. These are promising results to encourage further development of the DCNN-model and also better development of the synthetic data, both positive (FR) and negative (NFR).\\

\subsection{Random Validation Dataset}
    \label{s-random-validation-data}

The DCNN-models were developed and the training pathway optimized based on performance against the thirty-two events from the subset catalog, which are nicely behaved events.  The motivation for this was to have a fixed validation set that could be used for deeper analysis on DCNN-model performance and that was similar to the simulated training data. This left the remaining 321 events used in testing with relatively more edge and complicated cases. \\

With the in-depth analysis complete, and to see if a "wilder" validation set could steer the training process towards an improvement in performance, we ran the experiment again with a randomized selection of real events used in the validation step.  A set of 32 events, 16 of type E and 16 drawn from the F, F$^-$, and F$^+$ categories were randomly selected at the start of the experiment.  This random validation dataset was then used to pick the best model of each step-wise training process.  Table~\ref{T-random-eval} shows the results from this newly trained DCNN-model compared with the results from the previously trained DCNN-model. For comparison reasons the confusion matrix quantities for all 353 events (including Cx) are reported,  not just a part of it since the data sets were split differently.  All the metrics previous used (Accuracy, Precision, Recall and F1 Score) for the DCNN-model trained with a selected validation set (reference DCNN-model for this paper) are also reported for the DCNN-model steered with a random validation set. \\

We can see the results for the two different DCNN-models have very similar values for the No Noise and 5\% Noise case, within a tolerance of 3\% for no noise and 2\% for 5\% noise. The same doesn't happen for 10\% noise, with a difference of 10\% in precision, 0.1 in F1 Score and 0.2 in Recall, with the random validation set model having a better performance. Even though the DCNN-model using the random validation set has slightly better performance in general, it has lower number of TN and higher FP, showing a even stronger tendency to classify an event as FR. Even though there is some variation in the 10\% model, we can observe the same tendency across both DCNN-models when adding noise, bolstering the previous conclusion that we need a physically based fluctuation model to be implemented in the synthetic data. \\

\begin{table}[ht]
\centering
\caption{Comparison of classification metrics on all 353 events in the reference catalog when step-wise training was steered with a randomized validation set versus the fixed, simple validation set. Performance at three levels of noise: noise-free, 5\%, and 10\% noise are shown for each of the two experiments.}
 \begin{tabular}{@{\extracolsep{\fill}}ccccccc}\hline \hline 
  &  \multicolumn{3}{c}{{\centering Random Validation Set}} & \multicolumn{3}{c}{{\centering Selected Validation Set}}  \\
  \cmidrule(lr){2-4}\cmidrule(lr){5-7}
   & No Noise & 5\% Noise & 10\% Noise & No Noise & 5\% Noise & 10\% Noise \\
 \hline
TP & 117 (43\%)  & 217 (80\%)  &  198 (73\%) & 109 (40\%)  & 212 (78\%)  &  148 (54\%)\\
FP &  16 (20\%)  &  52 (64\%)  &   61 (75\%) & 14 (17\%)  & 50 (62\%)  &  45 (56\%)\\
TN  &  65 (80\%)  &  29 (36\%)  &   20 (25\%) & 67 (83\%)  & 31 (38\%)  &  36 (44\%)\\
FN & 155 (57\%)  &  55 (20\%)  &   74 (27\%) & 163 (60\%)  & 60 (22\%)  &  124 (46\%)\\
Accuracy & 52\%       &  70\%    &  62\%  & 50\%       &  69\%    &  52\% \\
Precision & 0.88 & 0.81 & 0.76  & 0.89 & 0.81 & 0.77\\
Recall & 0.43 & 0.80 & 0.73 & 0.40  & 0.78 & 0.54 \\
F1 Score & 0.58 & 0.80 & 0.75 & 0.55 & 0.79 & 0.64 \\
\hline  \hline
\end{tabular}
\label{T-random-eval}
\end{table}

In this paper we implement a non-traditional machine learning methodology that uses both synthetic data and some real data for training purposes. Some variation in the results, depending on which validation dataset and specific stopping criteria we use, is expected. The method to choose the validation dataset can, and will be, enhanced. Also more metrics criteria can be used to choose the best epoch for the step training process.

\section{Summary \& Conclusions}
    \label{S-Summary}
In this paper, we establish the framework for a novel technique not only to advance our understanding of the internal structure of ICMEs but also to pave the way to improve forecasting activities. Starting with the complex analysis of ICMEs' internal structure by \cite{Nieves_2019}, we develop a deep convolutional neural network (DCNN) model to classify \textit{in situ} signatures similarly. Training a DCNN is a time-consuming and costly task that typically involves collecting and analyzing a large amount of data to use in supervised learning. To handle the lack of real-world, labeled data, we combined two analytical flux rope models extracted from physical principles \cite[\cite{Nieves_C_2016} and ][]{Nieves_E_2018}, to act as a source of training data. We rely upon the technique of domain randomization, in which parameters of the simulator — such as angles, radius, velocity, magnetic field — are varied to induce the DCNN to learn the essential characteristics and peculiarities of the object of interest, i.e., flux rope signatures \citep{Tremblay_2018}. The DCNN-model was validated by analyzing metrics of the classification of a subset of the real data. It is reasonable to think that a similar approach could be applied to heliophysics fields on sparse or episodic data, such as the prediction of flares or solar energetic particle events, if a suitable simulation model exists for the training data.\\

The DCNN-model was able to classify between $76\%$ (in the final phase) and $84\%$ (in the subset evaluation phase) when training data with 5\% noise data is used. Precision and F1 Score are 0.78 and 0.88, respectively, for the evaluation set. Precision improves to .95 and F1 Score holds approximately constant at 0.86 in the test phase. Also, during the test phase, the classification accuracy jumped from 42\% when trained with noise free data to 76\% when trained on data with 5\% noise. The classification accuracy remains high at 69\% when trained with 10\% noise.  The results demonstrate good classification quality by having important statistical metrics (F1 Score, etc.) and similar scores between the evaluation subset and extended catalog.\\

The results demonstrate that the approach works. We were able to identify flux rope signatures using a pre-established DCNN handwriting model trained with synthetic data with high accuracy in well-behaved events. We have analyzed the discrepancies between manual and machine-learning based classification, and it opened a discussion on whether some events should be reclassified and how the classification criteria could be improved.\\

Moreover, the analysis of the classification discrepancies reinforced that flux rope models, especially physics-based flux rope models, are needed to understand the internal structure of ICMEs. Developing more models and including more observed features, such expansion, curvature or distortion, to the models will generate better training data.

Also, more physics-based fluctuation models should be explored and incorporated into the synthetic data (built-in or not in the flux rope models) for more realistic model fitting.\\

Future research will explore the methodology to implement the statistical and physical-based fluctuations observed in the data; synthetic complex (Cx) the simulation of more complex structures, and increase the number of synthetic events by changing the impact parameter. Once a satisfactory flux rope classifier is obtained, we will extend the DCNN-model to predict the best fitting parameters for each event.

\begin{acks}[Acknowledgments]
We thank Dr. Barbara Thompson and Tiago Pinho Da Silva M.S. for all discussions and reviews during the work done in this paper. We thank Marta Florido-Llinas BSA for making the flux rope interactive tool available.
Resources supporting this work were provided by the NASA High-End Computing (HEC) Program through the NASA Center for Climate Simulation (NCCS) at Goddard Space Flight Center. This material is based upon work supported by the National Science Foundation under Grant No. AGS-1433086. T. N-C also acknowledges to Goddard Strategic Collaboration Initiative. We acknowledge the tools used in this work. We used CUDA for processing (cuDNN) \citep{cudnn}, for data analysis and processing we used Numpy \citep{numpy}, Pandas \citep{pandas}, SciPy \citep{scipy}, and scikit-learn\citep{scikit-learn} and finally all plots were done using Matplotlib \citep{matplotlib}.
\end{acks}

\bibliographystyle{spr-mp-sola}
\bibliography{main}  

\appendix

\section{Metrics}
\label{A-Metrics}

Precision (Equation \ref{E-precision}) is the fraction of relevant instances among the retrieved instances, while Recall (Equation \ref{E-recall}) is the fraction of the total amount of relevant instances that were retrieved. Both Precision and Recall need to be taken into account when evaluating the performance of a predictive model. F1 Score (Equation \ref{E-fone}) is a well-established measure of a predictor's accuracy that considers both Precision (Equation \ref{E-precision}) and Recall (Equation \ref{E-recall}). Its ideal value is one, and worst value is zero.\\

\begin{equation}
\label{E-precision}
Precision = \frac{True Positive}{True Positive + False Positive} 
\end{equation}

\begin{equation}
\label{E-recall}
Recall = \frac{True Positive}{True Positive + False Negative} 
\end{equation}

\begin{equation}
\label{E-fone}
F_{1} Score = \frac{2 * Precision * Recall}{Precision + Recall} 
\end{equation}

\section{Complete classification}
\label{A-complete-table}

Table \ref{T-complete-events} contains the results for the classification done in all the 353 cases and has the necessary information to compare the classification done in the reference catalog and the classification done for the DCNN-model with different amounts of noise. The events marked with ``*" were used in the evaluation subset part of the training.

\begin{small}
\begin{longtable}{@{\extracolsep{\fill}}c|c|c|cccc}
    \caption{List of all the ICMEs from \cite{Nieves_2019}. For each selected ICME the table presents, event number, ICME Start Date, label assigned in by the reference catalog, classification with the no-noise model, classification with 5\% noise model, and classification with 10\% noise model. The events marked with ``*" were used in the evaluation part of the training.}\\

 \hline \hline 
 E & ICME Start & Catalog  & & Predicted Label & \\
   &      & Label               &  No Noise  & 5\% Noise & 10\% Noise        \\ 
 \hline\endfirsthead
 \hline \hline 
 E & ICME Start & Catalog  & & Predicted Label & \\
   &      & Label               &  No Noise  & 5\% Noise & 10\% Noise \\
\hline  \endhead
1&1995-02-07    &C$_x$  &NFR & FR &FR\\
2&1995-03-04    &F$_r$  &NFR & FR &FR\\
3&1995-03-06    &F$^-$  &NFR &NFR&NFR\\
4&1995-04-03    &F$^+$  &NFR &NFR&NFR\\
5&1995-04-05    &F$_r$  &NFR & FR &FR\\
6&*1995-05-13*  &E      & FR &FR&NFR\\
7&1995-06-30    &F$_r$  &NFR & FR &FR\\
8&1995-08-22    &F$_r$  & FR &NFR &NFR\\
9&1995-09-26    &F$_r$  &NFR &NFR&FR\\
10&1995-10-18   &F$_r$  & FR &FR&FR\\
11&1995-12-15   &F$^-$  &NFR & FR &NFR\\
\hline
12&1996-02-15&F$^+$ &NFR & FR &FR\\
13&1996-04-04&F$_r$ & FR &NFR &FR\\
14&1996-05-16&F$^+$ &NFR & FR &FR\\
15&1996-05-27&F$_r$ &NFR & FR &FR\\
16&1996-07-01&F$_r$ &NFR & FR &FR\\
17&1996-07-02&F$^-$ &NFR & FR &FR\\
18&1996-08-07&F$_r$ &NFR & FR &NFR\\
19&1996-12-24&F$^+$ & FR &FR&FR\\
\hline
20&1997-01-10&F$^+$& FR &FR&FR\\
21&1997-02-09&F$^-$&NFR &NFR&NFR\\
22&1997-04-10&F$_r$&NFR & FR &FR\\
23&1997-04-21&F$^+$& FR &FR&FR\\
24&1997-05-15&F$^+$& FR &FR&FR\\
25&1997-05-16&F$_r$& FR &FR&FR\\
26&1997-05-26&F$_r$& FR &FR&FR\\
27&1997-06-08&F$_r$&NFR & FR &FR\\
28&1997-06-19&F$_r$&NFR &NFR&NFR\\
29&1997-07-15&F$^+$& FR &FR&FR\\
30&1997-08-03&F$_r$& FR &FR&FR\\
31&1997-08-17&F$_r$&NFR & FR &FR\\
32&1997-09-02&F$_r$&NFR & FR &FR\\
33&1997-09-18&F$^+$&NFR & FR &FR\\
34&1997-09-21&F$^+$&NFR &NFR&NFR\\
35&1997-10-01&F$_r$&NFR & FR &FR\\
36&1997-10-10&F$^+$& FR &FR&FR\\
37&1997-11-06&F$^+$&NFR & FR &FR\\
38&1997-11-22&F$^+$& FR &FR&FR\\
39&1997-12-10&C$_x$&NFR & FR &FR\\
40&1997-12-30&F$_r$&NFR & FR &FR\\
\hline
41&*1998-01-06* &F$^+$& FR &FR&FR\\
42&1998-01-08   &F$^-$&NFR & FR &FR\\
43&1998-01-09   &F$_r$&NFR & FR &FR\\
44&1998-01-21   &F$^-$&NFR & FR &FR\\
45&1998-01-28   &F$^+$&NFR &NFR&FR\\
46&1998-02-02   &F$^+$&NFR & FR &FR\\
47&1998-02-04   &F$^+$&FR& FR &FR\\
48&1998-02-17   &F$_r$&FR& FR &FR\\
49&1998-02-18   &F$_r$&NFR & FR &FR\\
50&1998-03-04   &F$^+$&FR& FR &FR\\
51&1998-03-06   &C$_x$& FR &FR&FR\\
52&1998-03-25   &F$_r$&NFR &NFR&NFR\\
53&1998-03-31   &F$_r$&NFR & FR &FR\\
54&*1998-04-01* &E    &NFR&NFR &NFR\\
55&1998-05-01   &F$_r$&FR& FR &FR\\
56&1998-05-04   &F$^-$&NFR &NFR&FR\\
57&1998-06-02   &F$_r$&NFR & FR &FR\\
58&1998-06-24   &F$^+$&FR& FR &FR\\
59&1998-07-10   &F$^+$&NFR &NFR&FR\\
60&*1998-08-10* &E    &NFR& FR &FR\\
61&*1998-08-19* &F$^+$&FR& FR &FR\\
62&1998-08-26   &E    &NFR&NFR &NFR\\
63&1998-09-23   &F$^-$&FR& FR &FR\\
64&*1998-09-24* &F$^+$&FR& FR &FR\\
65&*1998-10-02* &E    &NFR&NFR &FR\\
66&1998-10-18   &F$^+$&FR& FR &FR\\
67&1998-10-23   &F$^-$&NFR & FR &NFR\\
68&1998-11-08   &C$_x$&NFR& FR &FR\\
69&*1998-11-09* &F$^+$&FR& FR &FR\\
\hline
70&1999-01-22  &E    &NFR&NFR &NFR\\
71&1999-02-11  &F$_r$&NFR & FR &FR\\
72&*1999-02-18*&E    &NFR& FR &FR\\
73&1999-04-16  &F$_r$&FR& FR &FR\\
74&1999-04-21  &E    &NFR& FR &FR\\
75&1999-05-28  &C$_x$&NFR& FR &FR\\
76&1999-06-26  &F$_r$&NFR & FR &FR\\
77&1999-07-02  &F$_r$&NFR &NFR&NFR\\
78&1999-07-06  &C$_x$& FR &FR&FR\\
79&1999-07-30  &E    & FR &FR&FR\\
80&1999-08-06  &F$_r$&NFR & FR &NFR\\
81&*1999-08-09*&F$_r$&FR& FR &FR\\
82&1999-09-15  &E    &NFR& FR &NFR\\
83&1999-09-21  &F$_r$&NFR & FR &FR\\
84&*1999-09-22*&E    &NFR&NFR &FR\\
85&1999-10-21  &E    & FR &FR&FR\\
86&1999-11-13  &E    & FR &FR&FR\\
87&1999-12-12  &C$_x$&NFR& FR &FR\\
\hline  
88&2000-02-11   &F$_r$&NFR & FR &FR\\
89&*2000-02-14* &E    &NFR&NFR &FR\\
90&2000-02-20   &F$_r$&FR& FR &NFR\\
91&2000-03-01   &F$_r$&FR& FR &FR\\
92&2000-03-28   &E    &NFR&NFR &NFR\\
93&2000-05-07   &C$_x$& FR &FR&FR\\
94&2000-06-08   &E    &NFR& FR &FR\\
95&*2000-06-23* &E    &NFR&NFR &NFR\\
96&2000-07-01   &F$_r$&FR& FR &FR\\
97&2000-07-11   &F$_r$&NFR & FR &FR\\
98&2000-07-13   &F$_r$&FR& FR &FR\\
99&2000-07-15   &F$^-$&NFR & FR &NFR\\
100&2000-07-15  &F$^+$&FR& FR &FR\\
101&2000-07-19  &C$_x$&NFR&NFR &FR\\
102&2000-07-28  &F$^+$&NFR & FR &FR\\
103&2000-07-31  &F$_r$&NFR &NFR&FR\\
104&2000-08-10  &F$^+$&NFR & FR &FR\\
105&2000-08-11  &F$^+$&FR& FR &FR\\
106&2000-09-02  &F$_r$&NFR & FR &FR\\
107&2000-09-04  &C$_x$&NFR&NFR &FR\\
108&2000-09-06  &C$_x$&NFR&NFR &NFR\\
109&2000-09-17  &E    &NFR&NFR &FR\\
110&2000-10-03  &F$^+$&FR& FR &FR\\
111&2000-10-05  &F$_r$&NFR & FR &FR\\
112&2000-10-12  &F$_r$&FR& FR &FR\\
113&2000-10-28  &F$^-$&FR& FR &FR\\
114&2000-11-06  &F$_r$&NFR & FR &FR\\
115&*2000-11-10*&E    &NFR&NFR &FR\\
116&2000-11-11  &E    &NFR&NFR &FR\\
117&2000-11-26  &F$_r$&NFR & FR &FR\\
118&2000-12-03  &C$_x$&NFR&NFR &NFR\\
\hline
119&2001-01-23  &C$_x$&NFR& FR &FR\\
120&2001-03-04  &C$_x$&NFR& FR &FR\\
121&*2001-03-19*&F$_r$&FR& FR &NFR\\
122&2001-03-20  &F$_r$&FR& FR &NFR\\
123&2001-03-27  &C$_x$&NFR&NFR &NFR\\
124&2001-04-04  &F$^-$&NFR & FR &NFR\\
125&2001-04-11  &F$^-$&NFR & FR &FR\\
126&2001-04-13  &F$^-$&NFR & FR &NFR\\
127&*2001-04-21*&F$_r$&FR& FR &FR\\
128&2001-04-28  &C$_x$&NFR& FR &FR\\
129&2001-05-27  &C$_x$&NFR&NFR &NFR\\
130&2001-06-27  &C$_x$&NFR&NFR &FR\\
131&*2001-08-05*&E    &NFR&NFR &NFR\\
132&2001-08-17  &F$^-$&NFR & FR &NFR\\
133&2001-09-25  &F$^-$&FR& FR &NFR\\
134&2001-09-29  &C$_x$& FR &FR&NFR\\
135&2001-09-30  &F$^-$&NFR & FR &FR\\
136&2001-10-02  &F$_r$&FR& FR &FR\\
137&2001-10-21  &F$_r$&NFR &NFR&FR\\
138&2001-10-28  &E    &NFR& FR &FR\\
139&2001-10-31  &F$_r$&FR& FR &FR\\
140&2001-11-24  &F$_r$&NFR &NFR&NFR\\
141&*2001-12-29*&F$_r$&FR& FR &FR\\
142&2001-12-30  &F$^-$&NFR &NFR&NFR\\
\hline
143&2002-02-28&F$_r$&NFR & FR &FR\\
144&2002-03-18&F$_r$&NFR & FR &FR\\
145&2002-03-23&F$_r$&FR& FR &FR\\
146&2002-04-14&F$^-$&NFR &NFR&NFR\\
147&2002-04-17&F$^+$&FR& FR &FR\\
148&2002-04-19&F$_r$&NFR & FR &FR\\
149&2002-04-23&F$^-$&NFR &NFR&NFR\\
150&2002-05-10&F$_r$&NFR & FR &FR\\
151&2002-05-11&F$_r$&FR& FR &FR\\
152&2002-05-18&F$_r$&FR& FR &NFR\\
153&2002-05-20&C$_x$&NFR& FR &FR\\
154&2002-05-23&F$^-$&NFR &NFR&FR\\
155&2002-07-17&F$_r$&FR& FR &FR\\
156&2002-08-01&F$_r$&FR& FR &FR\\
157&2002-08-01&F$_r$&NFR & FR &NFR\\
158&2002-08-18&F$_r$&NFR &NFR&FR\\
159&2002-08-26&F$_r$&NFR &NFR&FR\\
160&2002-09-03&C$_x$& FR &FR&FR\\
161&2002-09-30&F$^+$&FR& FR &FR\\
162&2002-11-16&F$^-$&NFR &NFR&NFR\\
163&2002-12-21&F$_r$&FR& FR &FR\\
\hline  
164&2003-01-26  &F$_r$&NFR &NFR&NFR\\
165&2003-02-01  &F$_r$&NFR & FR &NFR\\
166&2003-03-20  &F$_r$&NFR & FR &NFR\\
167&2003-05-09  &E    &NFR& FR &FR\\
168&2003-06-16  &F$_r$&NFR & FR &FR\\
169&2003-08-04  &F$_r$&NFR & FR &FR\\
170&2003-10-21  &C$_x$&NFR&NFR &NFR\\
171&2003-10-29  &C$_x$&NFR& FR &FR\\
172&2003-10-30  &C$_x$&NFR& FR &FR\\
173&*2003-11-20*&F$_r$&FR& FR &FR\\
\hline
174&2004-01-09  &E    &NFR& FR &FR\\
175&2004-04-03  &F$^+$&FR& FR &FR\\
176&2004-07-22  &C$_x$&NFR& FR &FR\\
177&2004-07-24  &F$_r$&NFR & FR &FR\\
178&2004-07-25  &F$_r$&NFR & FR &FR\\
179&2004-07-26  &C$_x$& FR &FR&FR\\
180&2004-08-29  &F$_r$&FR& FR &FR\\
181&*2004-09-13*&E    &NFR& FR &NFR\\
182&2004-09-17  &F$_r$&FR& FR &FR\\
183&2004-11-07  &F$_r$&FR& FR &FR\\
184&*2004-11-09*&F$^+$&FR& FR &FR\\
185&2004-11-11  &F$_r$&FR& FR &FR\\
186&2004-12-10  &F$^-$&NFR & FR &FR\\
\hline
187&2005-01-07  &F$^+$&FR& FR &FR\\
188&2005-01-08  &F$_r$&FR& FR &FR\\
189&2005-01-16  &F$^+$&FR& FR &FR\\
190&2005-01-18  &F$^-$&NFR & FR &NFR\\
191&2005-01-21  &F$^-$&NFR & FR &FR\\
192&2005-02-16  &F$^-$&NFR &NFR&FR\\
193&2005-02-17  &E    &NFR& FR &FR\\
194&2005-02-20  &F$_r$&NFR &NFR&NFR\\
195&2005-05-15  &F$^+$&FR& FR &FR\\
196&2005-05-20  &F$^+$&NFR & FR &FR\\
197&2005-06-12  &F$^-$&NFR & FR &FR\\
198&*2005-06-14*&F$^+$&FR& FR &FR\\
199&2005-07-10  &C$_x$&NFR&NFR &FR\\
200&2005-07-17  &F$_r$&NFR & FR &FR\\
201&2005-08-10  &F$^-$&NFR &NFR&NFR\\
202&2005-10-31  &F$_r$&NFR & FR &FR\\
\hline
203&2006-02-05  &F$^+$&FR& FR &FR\\
204&*2006-04-13*&F$^+$&FR& FR &FR\\
205&2006-04-14  &F$^-$&NFR & FR &NFR\\
206&2006-06-14  &F$^-$&NFR &NFR&NFR\\
207&2006-07-09  &C$_x$&NFR& FR &NFR\\
208&2006-08-19  &C$_x$&NFR&NFR &NFR\\
209&2006-08-30  &C$_x$&NFR& FR &FR\\
210&2006-09-30  &F$^+$&FR& FR &FR\\
211&2006-11-01  &F$^-$&NFR &NFR&NFR\\
212&2006-11-18  &F$_r$&NFR &NFR&NFR\\
213&2006-11-29  &F$^+$&FR& FR &FR\\
214&2006-12-14  &F$^-$&NFR & FR &FR\\
215&2006-12-16  &F$^-$&NFR & FR &NFR\\
\hline  
216&2007-01-14  &F$_r$&FR& FR &FR\\
217&2007-01-15  &F$^-$&FR& FR &FR\\
218&2007-03-29  &C$_x$&NFR& FR &FR\\
219&2007-05-21  &F$_r$&FR& FR &FR\\
220&2007-06-08  &F$_r$&NFR & FR &NFR\\
221&*2007-11-19*&F$_r$&FR& FR &NFR\\
222&2007-12-25  &F$^-$&NFR & FR &NFR\\
\hline
223&2008-05-23&F$^+$&NFR & FR &FR\\
224&2008-09-03&F$^+$&NFR & FR &FR\\
225&2008-09-17&F$_r$&FR& FR &FR\\
226&2008-12-04&F$_r$&NFR &NFR&NFR\\
227&2008-12-17&F$_r$&FR& FR &NFR\\
\hline
228&2009-01-02&F$^-$&NFR & FR &FR\\
229&2009-01-26&E    & FR &FR&FR\\
230&2009-02-03&F$^+$&NFR & FR &FR\\
231&2009-03-11&F$^+$&FR& FR &FR\\
232&2009-04-05&F$^-$&NFR &NFR&NFR\\
233&2009-04-22&F$_r$&FR& FR &FR\\
234&2009-06-03&F$_r$&NFR & FR &FR\\
235&2009-06-27&F$^+$&FR& FR &FR\\
236&2009-07-21&F$_r$&FR& FR &FR\\
237&2009-09-10&F$_r$&NFR & FR &FR\\
238&2009-09-30&F$_r$&FR& FR &FR\\
239&2009-10-29&F$^+$&FR& FR &FR\\
240&2009-11-01&F$^-$&NFR &NFR&NFR\\
241&2009-11-14&F$_r$&FR& FR &NFR\\
242&2009-12-12&F$_r$&NFR & FR &FR\\
\hline
243&2010-01-01  &F$_r$&NFR & FR &FR\\
244&2010-02-07  &F$_r$&NFR & FR &FR\\
245&2010-03-23  &F$_r$&NFR & FR &FR\\
246&2010-04-05  &F$_r$&FR& FR &FR\\
247&2010-04-11  &F$_r$&NFR & FR &FR\\
248&2010-05-18  &F$_r$&NFR & FR &FR\\
249&*2010-05-28*&F$_r$&FR& FR &FR\\
250&2010-06-21  &F$_r$&NFR & FR &NFR\\
251&2010-08-03  &C$_x$& FR &FR&FR\\
252&2010-09-15  &F$_r$&NFR &NFR&NFR\\
253&2010-09-25  &F$^-$&NFR &NFR&NFR\\
254&2010-10-11  &F$^-$&NFR & FR &FR\\
255&2010-10-31  &F$_r$&NFR &NFR&NFR\\
256&2010-12-19  &F$^+$&FR& FR &FR\\
\hline
257&*2011-01-24*&F$^+$&NFR & FR &FR\\
258&2011-02-18  &F$^-$&NFR &NFR&FR\\
259&2011-03-29  &F$_r$&NFR & FR &FR\\
260&2011-04-23  &F$^-$&NFR &NFR&FR\\
261&*2011-04-29*&E    &NFR&NFR &NFR\\
262&2011-05-28  &F$^+$&FR& FR &FR\\
263&2011-06-04  &F$_r$&NFR & FR &FR\\
264&2011-06-17  &C$_x$&NFR&NFR &FR\\
265&2011-06-30  &F$^-$&NFR & FR &NFR\\
266&2011-07-03  &F$_r$&NFR &NFR&NFR\\
267&2011-09-17  &F$_r$&NFR & FR &FR\\
268&2011-10-05  &C$_x$&NFR& FR &FR\\
269&2011-10-24  &C$_x$&NFR& FR &FR\\
270&2011-11-01  &F$^-$&NFR &NFR&NFR\\
271&2011-11-02  &F$_r$&FR& FR &NFR\\
272&2011-11-04  &C$_x$&NFR& FR &FR\\
273&2011-11-07  &F$_r$&NFR &NFR&NFR\\
274&2011-11-11  &C$_x$&NFR& FR &FR\\
275&2011-11-28  &C$_x$& FR &FR&FR\\
\hline
276&2012-01-21&F$_r$&FR& FR &FR\\
277&2012-01-22&F$^-$&NFR & FR &NFR\\
278&2012-02-14&F$_r$&FR& FR &FR\\
279&2012-02-26&C$_x$&NFR&NFR &NFR\\
280&2012-03-08&C$_x$&NFR& FR &FR\\
281&2012-03-12&C$_x$&NFR&NFR &FR\\
282&2012-03-15&F$_r$&FR& FR &FR\\
283&2012-04-05&F$_r$&FR& FR &FR\\
284&2012-04-11&C$_x$&NFR& FR &FR\\
285&2012-04-23&F$^-$&FR&NFR &NFR\\
286&2012-05-03&F$_r$&NFR &NFR&NFR\\
287&2012-05-16&F$_r$&NFR & FR &FR\\
288&2012-06-11&F$_r$&NFR & FR &NFR\\
289&2012-06-16&F$^+$&FR& FR &FR\\
290&2012-07-08&C$_x$&NFR& FR &FR\\
291&2012-07-14&F$_r$&FR& FR &FR\\
292&2012-08-12&F$_r$&NFR & FR &FR\\
293&2012-08-18&F$_r$&NFR &NFR&FR\\
294&2012-08-30&F$^-$&FR& FR &NFR\\
295&2012-09-01&F$_r$&NFR &NFR&FR\\
296&2012-09-04&F$_r$&NFR & FR &NFR\\
297&2012-09-06&F$^-$&FR& FR &NFR\\
298&2012-09-12&F$^-$&NFR &NFR&FR\\
299&2012-09-30&C$_x$&NFR& FR &FR\\
300&2012-10-08&F$_r$&FR& FR &FR\\
301&2012-10-12&F$_r$&FR& FR &NFR\\
302&2012-10-31&F$^+$&FR& FR &FR\\
303&2012-11-12&F$^+$&FR& FR &FR\\
304&2012-11-23&F$^-$&NFR &NFR&NFR\\
\hline
305&2013-01-16  &F$_r$&FR& FR &FR\\
306&2013-01-18  &F$_r$&NFR & FR &FR\\
307&2013-01-19  &F$^-$&NFR &NFR&NFR\\
308&2013-03-17  &F$_r$&NFR &NFR&NFR\\
309&2013-04-13  &F$^+$&NFR & FR &FR\\
310&2013-04-30  &F$_r$&FR& FR &NFR\\
311&2013-05-14  &F$_r$&FR& FR &FR\\
312&2013-06-06  &F$^+$&NFR & FR &FR\\
313&*2013-06-27*&F$_r$&FR& FR &FR\\
314&2013-07-04  &C$_x$&NFR& FR &FR\\
315&2013-07-12  &C$_x$&NFR& FR &FR\\
316&2013-09-01  &F$_r$&NFR & FR &FR\\
317&*2013-10-02*&E    & FR &FR&FR\\
318&2013-10-03  &F$_r$&NFR &NFR&NFR\\
319&2013-10-30  &F$_r$&FR& FR &FR\\
320&2013-11-08  &F$_r$&NFR & FR &NFR\\
321&2013-11-23  &F$_r$&NFR & FR &FR\\
322&2013-11-30  &C$_x$&NFR&NFR &FR\\
323&2013-12-08  &F$^-$&NFR & FR &FR\\
324&2013-12-14  &F$_r$&FR& FR &FR\\
325&2013-12-24  &F$^+$&FR& FR &FR\\
\hline
326&*2014-02-05*&E    &NFR&NFR &NFR\\
327&2014-02-15  &F$_r$&NFR & FR &FR\\
328&2014-02-18  &F$_r$&FR& FR &FR\\
329&2014-02-19  &C$_x$& FR &FR&FR\\
330&2014-04-05  &F$_r$&NFR & FR &FR\\
331&*2014-04-11*&F$^+$&FR& FR &FR\\
332&2014-04-20  &F$_r$&FR& FR &FR\\
333&2014-04-29  &F$_r$&FR& FR &NFR\\
334&2014-06-07  &C$_x$&NFR& FR &FR\\
335&2014-06-22  &F$^-$&FR& FR &NFR\\
336&2014-06-29  &F$_r$&NFR & FR &FR\\
337&2014-07-02  &F$^-$&NFR &NFR&NFR\\
338&2014-08-19  &F$^+$&FR& FR &FR\\
339&*2014-08-26*&F$_r$&NFR & FR &FR\\
340&2014-09-12  &F$^-$&FR& FR &FR\\
\hline
341&2015-01-07&F$^+$&FR& FR &FR\\
342&2015-03-28&F$_r$&NFR & FR &FR\\
343&2015-03-31&F$^-$&NFR &NFR&NFR\\
344&2015-04-09&C$_x$& FR &FR&FR\\
345&2015-05-06&F$^-$&FR& FR &FR\\
346&2015-05-08&F$^-$&NFR &NFR&NFR\\
347&2015-05-10&F$^+$&FR& FR &FR\\
348&2015-06-22&C$_x$&NFR&NFR &FR\\
349&2015-09-07&F$^+$&FR& FR &FR\\
350&2015-10-06&F$_r$&NFR &NFR&FR\\
351&2015-10-24&F$_r$&NFR &NFR&FR\\
352&2015-11-06&F$_r$&FR& FR &FR\\
353&2015-12-19&F$_r$&FR& FR &FR\\\hline  \hline
\label{T-complete-events}
\end{longtable}
\end{small}

\end{article} 

\end{document}